\documentclass[12pt,aps]{revtex4}
\usepackage{epsfig,amsmath,amssymb}

\newcommand{\bea}{\begin{eqnarray}}
\newcommand{\eea}{\end{eqnarray}}

\begin{document}

\vspace{1cm}

\title{Building the Standard Model \\
 Historical and Qualitative Aspects}

\vspace{0.5cm}

\author{Jan O. Eeg}

\email{j.o.eeg@fys.uio.no}
\affiliation{Department of Physics, University of Oslo,
P.O.Box 1048 Blindern, N-0316 Oslo, Norway}

\vspace{0.5 cm}

\begin{abstract}
  Without going to the details, I give a short, qualitative and historical
  description of  the development of
the Standard Model, from quantum electrodynamics,   
further  through   quantum chromodynamics, then to weak interactions with 
 parity- and CP-
violation, ending up with 
electroweak symmetry breaking and the Higgs boson. 

 The presentation is based on my own experiences from the late
 1960's up to the discovery of the Higgs boson. It is mainly meant for 
students at master level, and at some points it is presented 
somewhat different from standard 
textbook presentations.

\end{abstract}

\maketitle

\vspace{1cm}

\section{Introduction}

In the following  I give a very personal view on how the Standard Model 
for elementary particles appeared,
 step by step, studying quantum electrodynamics (QED), strong interactions
 and weak interactions. I describe   how I experienced it,
 first as a student until 
I am now professor  emeritus. Or in other words,  I tell how I saw
the growth of the Standard Model  from ca. 1967  to ca. 2012.

When I had finished courses in classical physics and standard non-relativistic
quantum mechanics, I learned  the relativistic Dirac equation, giving
for example relativistic
 corrections to the energy levels in Hydrogen. Then I studied
quantum electrodynamics(QED).  I liked  QED which I found to be a well
 formulated, consistent and interesting theory,
in contrast to some very phenomenological models for strong interactions.

\section{Quantum electrodynamics (QED)}

\begin{figure}
\begin{center}
\scalebox{0.3}{\includegraphics{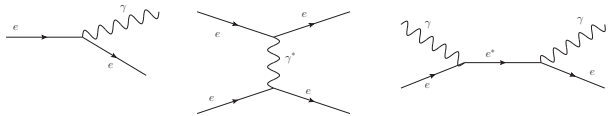}}
\caption{QED illustrated in terms of Feynman diagrams. The  diagram to the left
 is the building block, and the two other represent second order processes.
 The diagram in the middle shows scattering of two electrons(positrons). 
 The  diagram to the right describes Compton scattering, or annihilation
 of an electron-
positron pair to two photons, or the creation of an electron-positron pair
 from two photons, depending on which direction the diagram is read}
\label{basicQED}
\end{center}
\end{figure}

When I entered quantum  physics beyond non-relativistic quantum mechanics 
in 1967,
 QED was the only successful quantum field theory(QFT).
QED is  usually illustrated by {\it Feynman diagrams}, named after 
Richard Feynman who were one of the  three persons who got  the
nobel price for developing QED \cite{Feynman:1949hz}
(The two others were Julian Schwinger and
Sin-Itiro Tomonaga).
Some simple examples are given in  Fig. \ref{basicQED}. Here
 the diagram to the left describes the basic 
interaction in QED, where the
 solid line describes an electron (later and  more general a fermion)
 and the wavy
 line the photon (more general a gauge boson, also called an
 interaction particle).
In other words, 
this simple diagram illustrates an electron (fermionic) 
current interacting with an electromagnetic field particle, namely the
 photon ($\gamma$).
The diagram in the middle describes electron- electron (M\o ller-) scattering, 
illustrated as a photon line mediataing two electron currents, when time 
is running from left to right. The photon line illustrates
the Coulomb interactions between two electrons.
If time is running upwards, the same diagram illustrates $e^+ \, e^-$ (Bhabha-)
scattering.
The diagram to the right describes $e \, \gamma$ (Compton) scattering, 
illustrated as 
an  electron  absorbing and emitting again a photon (read from left to right)
. If time is running
 upwards , the same diagram is illustrating an electron-positron pair
 annihilating 
to two photons,  $e^+ \, e^- \, \rightarrow \,  \gamma \, \gamma$.
For these diagrams there are well defined mathematical expressions.
These diagrams correspond to the most simple lowest order QED processes.
QED might be thought to be described as series of terms using the left 
diagram in Fig. \ref{basicQED} as building block. Various processes are then
illustrated with a certain number of building blocks tied together with
electron and photon propagators.
 
What is new in quantum field theory beyond  quantum mechanics
 is the {\it quantum fluctuations}, also known as  {\it loop effects},
 where particles are appearing and disappearing over small time scales.
QED is verified to many orders in perturbation theory.
In all the Feynman diagrams, we know  the ingoing particles
 (the particles in the initial state) , and the outgoing particles
(the particles in the final state). The internal particles, for instance
 in loops, are  pictures of our mathematical expressions in 
perturbative theory. It should be noted that internal particles are 
``off-shell'', or off the mass-shell, that means that the relation 
$p^2 = m^2$ is broken.(Here $p$ is the four momentum and
$m$ the mass of the  particle).

In  Quantum Mechanics  given by the Schr\"odinger equation   and also
 the relativistic  Dirac equation, there are  three important results
 which were found to be only approximatele correct when we proceed to
 the full QED.

\begin{itemize}

\item  First, the energy levels 
$E(2S_{1/2})$ and $ E(2P_{1/2})$ in Hydrogen were equal (degenerate). 

\item Second, the magnetic moment of electron was given by 
$\vec{\mu} \, = \, \frac{e g_e}{2 m_e} \vec{S}$ with $g_e =2$.

\item Third, the 
strength of electromagnetic interactions were given by
$\alpha_{em} \, \equiv  \, \frac{e^2}{2 \epsilon_0 h c} \;
\simeq \frac{1}{137} . $

\end{itemize}

{ \it These three results was shown to be slightly  changed
 in the full
 quantum electrodynamics(QED), due to quantum fluctuations
 as will be discussed below.}

Concerning the Dirac equation itself it has been introduced in many ways.
May favorite version was to introduce it as a Schr\"odinger type
equation with an 
Hamilton operator which contained the momentum operator 
$(\vec{p} = -i \hbar \vec{\nabla})$ to first order.
 This was a reasonable assumption for a relativistic equation because,
 written in Hamiltonian form, it is of first order in the time derivative.
 But then one also assumed that the Dirac equation contained some-a
 priori- unknown matrices.
 To obtain the Dirac equation one
 required  that the  squared Hamiltonian  is equal to 
 $(c \vec{p})^2 + (mc^2)^2$. Then one
 finds that the matrices must be anticommuting, and have the dimension 
 $4 \times 4$, minimally.

\subsection{ ``Feynman's approach to QED ``}

Feynman entered QED with his famous intiution and he  
solved the  equations without the complicated 
machinery with field operators.
As I understood it, 
he studied a priori differential equations for scattering of
 an electron in an electromagnetic field. 
Then these equations were 
adapted to more complicated processes. This method, often named
the ``Feynman approach'' is
nicely described
in the old book ``Relativistic quantum mechanics'' written by
Bjorken and Drell \cite{Bjorken:1965sts}. I learned, and later
lectured, some times this version which is
theoretically simplified 
compared to standard QED with all the operator formalism.
However, there is an important warning: Using this simplified version, 
the Pauli principle had to be put in by hand!,- in contrast to true QED
field theory, where all the correct signs are automatially obtained by
the anticommuting fields.
The well known  equations to be solved  are
\begin{equation}
 \left( \gamma_\mu  \,(i \partial^\mu \, -  \, e A^\mu(x)) 
- \, m \right) \, \Psi(x) \, = \, 0 \quad ; \; \; 
\partial_\nu \partial^\nu \, A^\mu(x) \, = \, e j_D^\mu(x) \; , 
\label{ClassEq}
\end{equation}
where $e$ is the elementary electron charge. Further,
$A^\mu(x)$ is the electromagnetic field  
and  $\Psi(x)$
is the electron field which would in the non-relativistic case  
correspond to the electron wave function.
Here  Lorentz gauge for the electromagnetic field is assumed: 
\begin{equation}
\partial_\nu A^\nu(x) \, = 0  \; .
\end{equation}
Now, for the electromagnetic current  in a scattering process one writes
the Dirac current as the {\it transition current} 
 \begin{equation}
 j_D^\mu(x) \, = \,
 \overline{\psi(x)}_b \gamma^\mu \psi(x)_a \; ,
\label{elmcurr}
\end{equation}
where $\psi(x)_a$ represents an ingoing (asymtotically) free electron
in some chosen process and  $\overline{\psi(x)}_b$
represents an outgoing (asymtotically) free electron in the same process.

\vspace{0.3cm}
The differential equation for $\Psi(x)$ in (\ref{ClassEq}) can be rewritten
\begin{equation}
 \left( \gamma_\mu  \,(i \partial^\mu \, -  \, e A^\mu(x)) 
- \, m \right) \, \Psi(x) \, =   \; 
e A^\mu(x) \gamma_\mu \Psi(x) \; ,
\label{ClassEq2}
\end{equation}

The equation in (\ref{ClassEq2}) for $\Psi$ and the equation
for $A^\mu$ in (\ref{ClassEq})
may be solved by free Green-functions $S_F(x-y)$  
for the free Dirac Field $\Psi(x)$ and $D_F(x-y)$ for the free vector
field  $A^\mu(x)$ respectively :
\begin{equation}
\left(i \gamma_\mu \partial^\mu \, - \, m \right)S_F(x-y) = 
{\bf 1}^{(D)}
\delta^{(4)}(x-y)  
\quad , \; \; \mbox{and} \; \, 
\partial_\nu \partial^\nu D_F(x-y)_{\alpha \beta}  =
\delta_{\alpha \beta}\delta^{(4)}(x-y),
\end{equation}
where ${\bf 1}^{(D)}$ is the unit matrix  in the space of Dirac matrices, and
$\mu, \nu, \alpha, \beta$ are Lorentz indices.
These Greens functions will (for proper boundary conditions )
also be the propagators for the particles.
They are given by
\begin{equation}
  S_F(x-y) = \int \frac{d^4 p}{(2\pi^4}
  \frac{(\gamma \cdot p \, + \, m)}{(p^2-m^2 +i \epsilon)} \; \, ,
  \mbox{and} 
  \qquad D_F(x-y)_{\alpha \beta} =  \int \frac{d^4 k}{(2\pi^4)}
  \frac{(-g_{\alpha \beta})}{(k^2 + i \epsilon)} \; \, .
\label{DirProp}
\end{equation}
  Here the small $\epsilon$ takes care of the asymtotic properties
  of the propagator (for $\epsilon \rightarrow 0$)
 For the electron(fermion)propagator  one also  obtains :
\begin{equation}
S_F(x-y) = \theta(x_0-y_0) \sum_{\sigma} \int\frac{d^3 p}{(2\pi)^3}
 \psi^{(+)}_{p \sigma}(x) \overline{\psi^{(+)}_{p \sigma}(y)} \,  - \,
\theta(y_0-x_0) \sum_{\sigma} \int\frac{d^3 p}{(2\pi)^3}
 \psi^{(-)}_{p \sigma}(x) \overline{\psi^{(-)}_{p \sigma}(y)} \; ,
\end{equation}
where $\psi^{(\pm)}_{p \sigma}(x)$ are plane wave solutions 
of the Dirac equation with
definite momentum ($p$)and spin quantum number $\sigma$,
and energy sign $\epsilon = \pm$.
$\psi^{(+)}$ and $ \psi^{(-)}$ represent the particle and anti-particle
solutions respectively.
As usual $\overline{\psi} = (psi)^dagger \, \gamma_0$.
 A similar expression for the photon propagator is cumbersome because of
 the gauge freedom for the photon field. The photon propagator is presented
 in various ways, for instance via {\it path integrals}.
 
For  the interacting electron field one obtains the solution:
\begin{equation}
\Psi(x)_i \, = \, \psi(x)_i \, + \, \int d^4 y \, S_F(x-y) \, 
e \gamma^\mu A_\mu(y) \, \Psi(y)_i \; \, ,
\end{equation}
where $\Psi(x)_i$ is the solution when  $\psi(x)_i$ is the ingoing plane wave.
As it stands, this solution looks useless because the unknown solution
$\Psi(x)_i$ is also under
 the integral at the right hand side. In order to obtain useful result
 one  solves the equation iteratively, or  in other words, perturbatively:
\begin{equation}
\Psi(x)_i \, = \, \sum_n \Psi(x)^{(n)}_i \; ,
\label{PertExp}
\end{equation}
where $\Psi(x)^{(0)}_i = \psi(x)_i$ and the $n$'th
 term is given by
\begin{equation}
\Psi(x)^{(n)}_i \, = \, \, \int d^4 y \, S_F(x-y) \, 
e \gamma^\mu A_\mu(y) \,\Psi(y)^{(n-1)}_i \; \, . 
\end{equation}
This gives the first non-trivial approximation
\begin{equation}
\Psi(x)^{(1)}_i \, = \, \, \int d^4 y \, S_F(x-y) \, 
e \gamma^\mu A_\mu(y) \,\psi(y)_i  \; \, .
\end{equation}
Further, the second term is 
\begin{equation}
\Psi(x)^{(2)}_i \, = \, \, \int \int d^4 y \,d^4 z S_F(x-y) \, 
e \gamma^\mu A_\mu(y) \, S_F(y-z) e \gamma^\nu A_\nu(z) \psi(z)_i \; \, , 
\end{equation}
and so on. 
\vspace{0.1cm}
The solution for the electromagnetic field is: 
\begin{equation}
  A^\mu(x) \, = \,A^\mu_0(x) \, + \,  \int d^4 y \, D_F(x-y)^{\mu \nu} \,
  e j_{\nu D }(y) \; \, ,
  \label{A-equation}
  \end{equation}
where $A^\mu_0(x)$ is the initial (some times the final) free solution
electromagnetic field, depending on the 
process which is considered, and $j_D^\mu(y)$ given in (\ref{elmcurr}).
When there are no photons in the initial or final state, then $A^\mu_0(x)$
is zero.

The scattering matrix is -for a given initial state $\psi_i$ - the probability
amplitude to measure a chosen final state $\psi_f$ to occur. Thus the 
 scattering matrix is given by a scalar product of the 
interacting  $\Psi_i$ and the chosen  final plane wave $\psi(x)_f$ :
\begin{equation}
 S_{f i} \, = \,  \left(\psi(x)_f , 
\Psi(x)_i\right)_{x_0 \rightarrow \epsilon_f \infty} = 
\, \delta_{fi} \, + \, \int d^4y \, e \,  
\overline{\psi(y)_{f}} \, \gamma^\mu A_\mu(y) \, \Psi(y)_i \; \, .
\label{Smatrix}
\end{equation}
Here $x_0 \rightarrow \, + \infty$ when $\psi_f$ is a particle 
(positive frequency $\epsilon_f$ = +1) and $x_0 \rightarrow \, - \infty$
 when $\psi_f$ is
 an antiparticle (negative frequency $\epsilon_f$ = -1). This means
 that mathematically, 
in outgoing(ingoing) particle with negative energy is intepreted as an
 ingoing(outgoing) antiparticle. 

The scattering matrix 
 can be written as a perturbation series, when $\Psi(y)_i$
is given by equation (\ref{PertExp}):
\begin{equation}
 S_{f i} \, = \,  \sum_n  \; S_{fi}^{(n)} \quad  \mbox{where} \; \; 
S_{fi}^{(0)} = \delta_{fi} \; \, ,
\label{SExp}
\end{equation}
and further, the first nontrivial term is 
\begin{equation}
 S_{fi}^{(1)} \, = \, \epsilon_f \, e \, \int d^4y  \, 
\overline{\psi(y)_{f}} \, \gamma^\mu A_\mu(y) \, \psi(y)_i \; \, .
\label{SExp1}
\end{equation}
Here we  obtain simple scattering of an electron in a Coulomb field
$A^\mu_0= A^{\mu}_{Coul}$ and put $j_D^\mu=0$ in (\ref{A-equation}). Or,
alternatively, we may put
$A^\mu_0 = 0$ and use a suitable Dirac current $j^\mu_D$ for an
in- and out-going
electron(positron); see equation (\ref{elmcurr}). Then we  describe
scattering of an electron (positron) in the field of another
electron(positron), that is, we may describe electron-electron (M\o ller)
scattering or electron-positron (Bhabha) scattering;- {\it provided we
symmetrise the following amplitude according to the Pauli principle}.
Mathematically, for a relevant contribution
\begin{equation}
 S_{fi}^{(1,j_D)} \, = \,  \epsilon_f  \, e \, \int  d^4y  \,
 \overline{\psi(y)_{f}} \, \gamma^\mu \psi(y)_i \,\int d^4x D_F(x-y)_{\mu \nu} \,
 [e \overline{\psi(x)_b} \gamma^\nu \psi(x)_a] \; \, .
\label{SExp1jD}
\end{equation}
In standard QED this would correspond to a second a second order contribution 
from the scattering operator contribution.

The next non-trivial term is 
\begin{equation}
 S_{fi}^{(2)} \, =   \, \epsilon_f  e^2 \, \int \int d^4y \, d^4z   \, 
\overline{\psi(y)_{f}} \, \gamma^\mu A_\mu(y) \,
\, S_F(y-z) e \gamma^\nu A_\nu(z) \psi(z)_i \; \, ,
\label{SExp2}
\end{equation}
which may describe Compton scattering, or
$e^+ \, e^- \rightarrow \gamma \gamma$
annihilation , or pair creationn $\gamma \gamma \rightarrow e^+ e^-$.

The method described above works, and gives the same result as the
QED with the fields
 as operators, (again!:)  provided that
the minus sign is used if the final electron is an antiparticle.
Antiparticles (positrons) are represented by negative enrgy electrons 
going backwards in time.
There is also the obscure intepretation of the  negative
energy solutions filling up the  ``Dirac sea''.
In  QED with anticommuting fermion field operators the correct relative signs
come out automatic, and the obscure Dirac sea disappears.
Note that (\ref{SExp2}) has the same form as the second order expression 
in QED where the $\psi$'s are field operators, except for the factor $1/(2!)$.
However such  statistical factors typically disappears (replaced by the number one) when Wicks theorem are applied for the field operators.
Some lecturers do not mention this
``relativistic quantum mechanics'' (``Feynman approach'') because of its
shortcomings. Others thinks it is OK to see how close one might come with
some intiution.

\subsection{ Renormalizationn in QED}

The QED Lagrangian may be written
\begin{equation}
{\cal L}_{QED} \, = \, 
\label{LQED}
\end{equation}
\begin{equation}
\overline{\psi(x)} \left( i \gamma_\mu  \, \partial^\mu \, 
- \, m \right) \, \psi(x) \hspace{1cm}
\rightarrow (\mbox{electron propagator})     
\label{ElProp}
\end{equation}
\begin{equation}
 - \frac{1}{4} F_{\mu \nu}(x) \, F^{\mu \nu}(x) \hspace{2.5cm} 
 \rightarrow (\mbox{photon propagator})     
\label{PhProp}
\end{equation}
\begin{equation}
- i e A_\mu(x) \, \left(\overline{\psi(x)} \gamma_\mu \psi(x) \right)
\hspace{1,3cm} \rightarrow (\mbox{interaction}) \; ,
\label{IntTerm}
\end{equation}
\vspace{0.1cm}
where $F_{\mu \nu}$ is, as described in some textbooks,
given by the commutator of two covariant
derivatives $iD_\lambda = \partial_\lambda \, \, -   \, e A_\lambda$:
\begin{equation}
i e \, F_{\mu \nu}(x) \; = \; [iD_\mu \, , \, iD_\nu] \, = \, 
ie (\partial_\mu A_\nu \, - \, \partial_\nu A_\mu)
\end{equation}
Using standard canonical quantisation, from this  Lagrangian 
one may calculate all processe in QED.
But one should note some  important points.:

In relativistic physics, particles may appear and disappear
as quantum fluctuations (loop effects) as long as the two quantities
{\it energy  and electric charge are conserved!}

Further,  {\it a free electron should be stable.} This corsponds to the
 $\delta_{f i} \, = \, 1$ with
$f=i$. This means that the integral in (\ref{Smatrix}) should not contribute
 to the S-matrix for a free electron. However, a concrete calculation
 shows that this requirement is not fulfilled.
 This means that  for the electron part one obtains
 a factor different from one  in front of the first term in (\ref{ElProp}).
 To solve this problem, one has to realize that the fields in
 the Lagrangian above are {\it bare fields}. The physical {\it renormalized 
} fields, dressed with loop effects are $\Psi_R =  \sqrt{Z_F} \Psi_0$, 
 where $\Psi_0$ is the bare field in the Lagrangian above, and $Z_F$
 represent the loop effects.
 Also the mass term in (\ref{ElProp}) has to be
 renormalised ($m_0 \rightarrow m_R$).
 And also the photon term (\ref{PhProp}) and the 
interaction term (\ref{IntTerm}) has also to be renormalized. 
This will be discussed in the following  subsections.

\subsection{ Lowest order loop effects}

\begin{figure}
\begin{center}
\scalebox{0.6}{\includegraphics{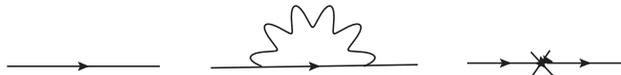}}
\caption{Self-energy correction(s) to electron propagator}
\label{basicQEDA}
\end{center}
\end{figure}

In Fig. 1 the left diagram (just the electron line) corresponds to the 
bare electron propagator.
The lowest order correction to the electron propagator (\ref{ElProp})
 contribution to the electron paropagator, is given
 in the middle. As already mentioned, the electron is stable,
and the theory(QED) should be constistent with this fact.
 Therefore 
the $S$-matrix element corresponding 
to the  to the diagram in Fig. \ref{basicQEDA} (lowest 
order self energy)  should be zero for an electron on the physical mass-shell 
$p^2 = m^2$.(Also for $\gamma \cdot p \rightarrow m)$.
 A concrete calculation shows that this is not the case. In order to
 solve this problem , one has to redefine the mass and  write
\begin{equation}
m_R \; \, = \; \, m_0 \, -  \, \delta m \; \, ,
\end{equation}
where $m_0$ is the {\it bare mass}  contained in (\ref{ElProp}) and $\delta m$ is the result from the diagram in the middle of Fig. \ref{basicQEDA}. 
 Now the right diagram in 
 Fig. \ref{basicQEDA}  (the {\it counterterm}) 
with the cross to the
 right substracts the value of diagram 1b to get 
 zero on-shell $p^2 = m_R^2$.
An a priori disturbing  discovery was also that $\delta m$ was infinite,
 and could be parametrised
 through a cut-off $\Lambda$ :
\begin{equation}
\delta m \, = m_0 \, \frac{\alpha}{4 \pi} ln(\frac{\Lambda^2}{m_0^2})
\, .
\end{equation}

Moreover, it can be shown that in order to get the propagator
pole at $m_R$, the electron part has to be multiplied by a
additional factor
$Z_F = 1 + k \frac{\alpha}{4 \pi} ln(\frac{\Lambda^2}{m_0^2})$
where $k$ is some number of order one given by the loop integration.
Then also the pole of the electron propagator will be shifted
from $p^2 =m_0^2$ in (\ref{DirProp} )
to the
 correct place, $p^2 = m_R^2$.  Still the remaining part of diagram 
in Fig. \ref{basicQEDA}
  for off-shell electrons are nonzero!
It turns out that for a bound electron (which is off-shell)
 in a Hydrogen atom 
this non-zero off-shell part will give the main contribution
 to Lamb-shift.
Namely:
  The degenerations of  $2S_{1/2}$  and $2P_{1/2}$ in Hydrogen 
are now removed (lifted) by the (renormalized) self-energy contribution for 
{\it bound} (off-shell) electrons. 
This gives the famous Lamb-shift in Hydrogen  giving the 
 $\lambda \simeq $ 21 cm   line.

\begin{figure}
\begin{center}
\scalebox{0.4}{\includegraphics{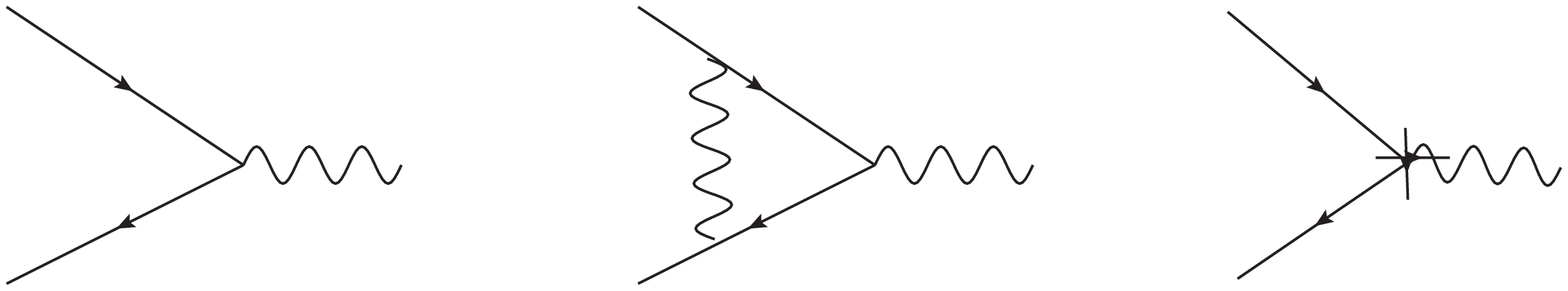}}
\caption{Vertex correction(s) for electromagnetic  interaction}
\label{basicQEDB}
\end{center}
\end{figure}

Within non-relativistic quantum mechanics, and also
for the relativistic equation of Dirac, the gyromagnetic 
factor of the electron is $g_e =  2$. Entering QED, there are 
corrections to this result.
Now the diagram in the middle of Fig. \ref{basicQEDB} 
  will give a contribution(correction) to
the gyromagnetic factor.
Also here there is a diagram (a counterterm) to the right of 
Fig.\ref{basicQEDB} 
 to subtract unphysical terms via
renormalisation. 
Now  the remaining vertex correction
( the middle 
of Fig.  \ref{basicQEDB} ) gives a small correction to 
$g_e$ such that with the lowest order correction, the result is 
\begin{equation}
g_e \, = \,  2 + \frac{\alpha_{em}}{\pi}
\end{equation}

\subsection{ Vacuum polarisation}

\begin{figure}
\begin{center}
\scalebox{0.5}{\includegraphics{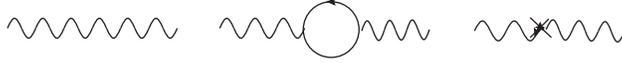}}
\caption{Diagrams for the  photon propagator. To the left is the bare
  propagator. In the middle is the  diagram which
  gives the vacuum polarisation effect. To the right is the corresponding
  counter term. }
\label{vacPol}
\end{center}
\end{figure}

In Fig \ref{vacPol} the left diagram, just the wavy line, represents 
the bare photon propagator in (\ref{PhProp}).
The diagram in the middle of Fig \ref{vacPol} gives a corrections to 
the photon propagator, usually
 called the {\it vacuum polarisation diagram}.
Adding the conterterm in 4c will stabilise the photon.
The diagram in the middle of Fig. \ref{vacPol} means that 
electron-positron pairs appear and disappears
(as {\it quantum fluctuations}). This can be seen as 
screening of electric charge, which  gives an energy dependent  fine structure 
``constant'' $\alpha_{em}$.
Previously this diagram(s) were interpreted as a correction to the 
Coulomb potential.
Now, in modern field theory, one usually intepretes this as 
an energy dependent effective fine structure. This means that
   $\alpha_{em} \, = \, \frac{1}{137}$ for low energy photons.   
 The momentum-energy  dependent fine structure factor is 
$\alpha_{em}(q^2) \, > \, \frac{1}{137}$
for larger energies.
Also this means thet the electric charge in eq. (\ref{IntTerm}) is the bare
 charge $e_0$. 
Adding later also other charged  particles in the loop, i.e more
 charged leptons,
 quarks and also the charged W-bosons,
 make  $\alpha_{em} \, \simeq \, \frac{1}{128}$ 
at the energy $M_Z c^2$.

\section{Strong interactions ?}

In the 1960's, 
weak and strong interactions were poorly understood theoretically.
People worked on the  ``Regge poles''
 for strong interactions \cite{Regge:1953fz}. 
Such models were  popular in the 1960s and 1970's.
In this case one cosidered the scattering amplitude, and studied
  poles and trajectores in a complex plane.

 Also some theoreticians worked on formal S-matrix theory, where scattering
 amplitudes were  considered
 to be maximally analytical functions  in some complex plane. Thresholds for
 scattering gave singularities in the complex plane.

 \subsection{ Pion-nucleon interactions }

One tried to construct a theory for interactions between pions and nucleons
as a field theory where photons were replaced by pions and electrons
 by nucleons in an isospin symmetric way.
\begin{figure}
\begin{center}
\scalebox{0.4}{\includegraphics{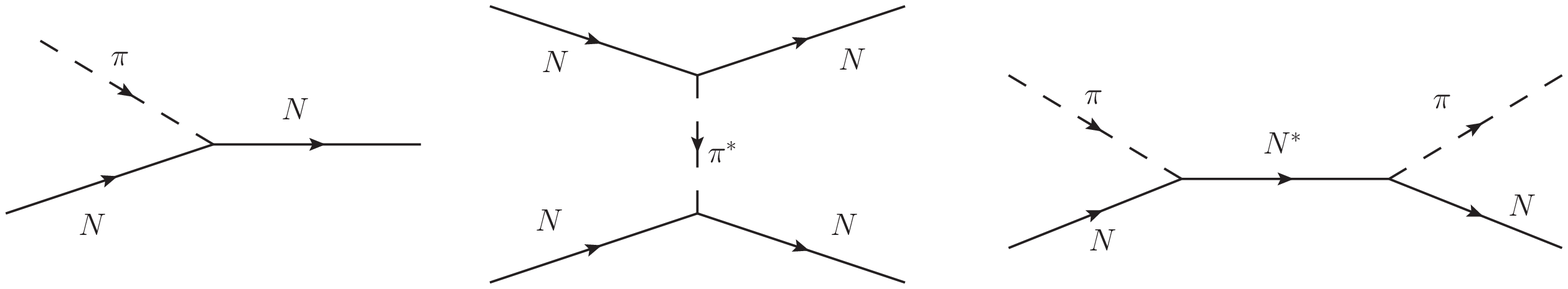}}
\caption{Some $\pi$-Nucleon Feynman diagrams
  }
\label{piN-diagrams}
\end{center} 
\end{figure}
The  isospin symmetric interaction Lagrangian for  strong interactions:
was written down as
\begin{eqnarray}
{\cal L}_{\pi N} \; = \; G_{\pi N} \, \bar{N} \, (\tau_i \, \Phi_i) \,
  \gamma_5 \, N  \; \; , \;  N \; = 
  \;  \left( \begin{array}{c} p \\
 n  \end{array} \right) \; \, , 
\label{Lstrong}
\end{eqnarray}
where $N$ is the nucleon doublet, $\Phi$ the pion triplet
and  $\tau$ the Pauli matrices which here represents isospin.
Here 
\begin{eqnarray}
  \pi^{\pm} \, = \, \frac{1}{\sqrt{2}}(\Phi^1 \, \pm i \Phi^2) \; \, , \;
  \pi^0 \, = \, \Phi^3 \; \, ,
\label{pis}
\end{eqnarray}
are the fields for physical charged and neutral pions.
Inspired by the sucess of QED,
one hoped to write some QED-like theory for
nucleons and pions. In addition one incorporated isospin
 symmetry for pions and nucleons.
This was in some books presented as a fruitful example for a field theory.
From the pure mathematical point of view this theory was OK. BUT: Taking
into account experimental results for pi-nucleon scattering,
the pi-nucleon coupling $G_{\pi N}$ had to be much
too big for a perturbative expansion (in contrast to QED):
\begin{equation}
\frac{(G_{\pi N})^2}{4 \pi} \; \, \simeq \; \, 15  \; ,
\label{Gsquare}
\end{equation}
making perturbative theory useless.

\subsection{The structure of hadrons -  $SU(3)_F$ flavor symmetry }

\begin{figure}
\begin{center}
\scalebox{0.35
}{\includegraphics{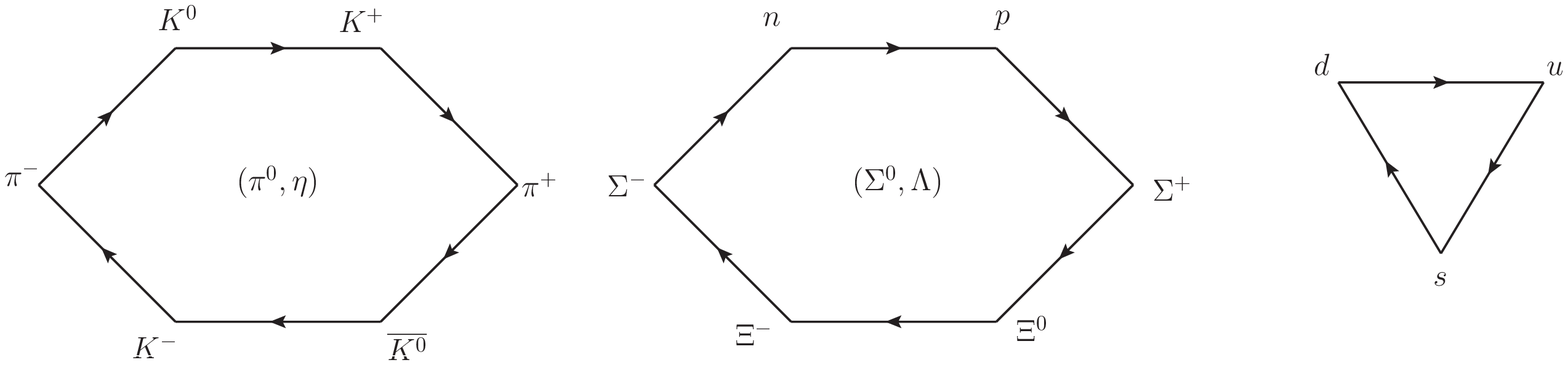}}
\caption{Pseudoscalars ($0^-$, left figure) and
 baryons$(\frac{1}{2}^-)$(figure in middle)  put in  octets (8-plets) 
representations of $SU(3)$. This was making order i hadronic particles.
 BUT the elementary triplet (3-plet) 
representation (``quarks'') {\it was apparently not realized - or?} 
 }
\label{FlavorDiagram}
\end{center}
\end{figure}

It was shown by Gell-Mann \cite{Gell-Mann:1961omu} that the
 pions and kaons make an octet of pseudo-scalar particles ($0^-$)
within $SU(3)$-symmetry, as illustrated in Fig.\ref{FlavorDiagram}.
There were also a similar baryon octet containining the proton, neutron and 
the strange $\Sigma$ and $\Xi$-particles.

More particles were found, for instance the were spin 1-octets containing 
the $\rho,K^*,\omega,\phi$-particles, and a $SU(3)$ decuplet containing
the spin 3/2 particles $\Delta,\Sigma^*,\Xi^*,\Omega^-$.
All these particles were explained as built up of the fundamental
$SU(3)$- triplet $(u,d,s)$. There was also the corresponding anti-triplet
($\bar{u},\bar{d},\bar{s}$).
The $u$ should then have charge +2/3 and both $d$ and $s$ have
charge -1/3 of the proton electric charge.

Examples: 

\begin{itemize}

\item
  Pseudoscalars:  $\pi^+ =(u \bar{d}), \; \,
  \pi^0 =(u \bar{u} - u \bar{u})/\sqrt{2}, \; \,  
  \pi^- =(d \bar{u}), \; \,  K^+ =(u, \bar{s}), \; \,  K^0 =(d, \bar{s}), \; \,
  K^- = (s, \bar{u}), \; \, 
\overline{K^0} = (s,\bar{d}), \; \eta_8 = 
(u \bar{u} + d \bar{d} - 2 s \bar{s})/\sqrt{6} $

\item
  Spin 1/2 baryons : $p = (uud), \; \, n =(ddu), \; \, \Sigma^+ =(uus), \; \,
  \Sigma^0 = (uds), \; \, 
  \Sigma^- =(dds), \; \,  \Lambda =(uds), \; \, \Xi^0 = (uss), \; \,
  \Xi^- = (dss)$

\item
  Spin one  (Vectors):  $\rho^+ =(u \bar{d}), \; \, $
  $\rho^0 =(u \bar{u} - u \bar{u})/\sqrt{2}, \; \,
  \rho^- =(d \bar{u}), \; \,$
  $K^{+ *} =(u, \bar{s}), \; \, $ $ K^{0 *} =(d, \bar{s}), \; \, $
  $K^{- *} = (s, \bar{u}), \; \, $
 $\overline{K^{0 *}} = (s,\bar{d}), \; \, $
  $\omega =  (u \bar{u} + d \bar{d})/\sqrt{2} , \; \, $
  $\Phi= (s \bar{s})$

\item
  Spin 3/2 baryons : $\Delta^{++} = (uuu), \; \, $ $\Delta^+ = (uud), \; \, $
  $ \Delta^0 =(ddu), \; \, $
  $\Delta^- =(ddd), \; \, $
  $\Sigma^{*+} =(uus), \; \, $ $ \Sigma^{*0} = (uds), \; \, $
  $\Sigma^{*-} =(dds), \; \, $  $\Xi^{*0} = (uss), \; \,$
  $\Xi^{*-} = (dss) , \; \, $ $ \Omega^- = (sss)$

\end{itemize}

BUT: The fundamental triplet $(u,d,s)$ itself were apparenly absent.
It was in the beginning thought to be an {\it auxilary object}!

The electric charges of the particles within $SU(3)_F$ flavor symmetry
can be expressed as
\begin{equation}
  Q \, = \, I_3 \, + \, \frac{Y}{2} \; , 
  \label{chargeForm}
\end{equation}
where $I_3$ is the isospin measured along the horisontal axis,
and $Y$ is the hypercharge measured along the vertical axis in the
diagram in Fig \ref{FlavorDiagram}.
As more and more baryons and mesons were discovered, some physicists 
were starting to think that all these mesons and baryons
were not elementary particles,
but were composed from ``smaller'', more fundamental particles.

In Fig. \ref{FlavorDiagram} we see the $0^-$ meson and $(1/2)^+$
baryon $SU(3)$ octets.
 Maybe the elementary
 triplet could correspond to fundamental constituents of the baryons 
and mesons?  Such fundamental particles were called {\it quarks} or simply
{\it partons.}. This hypothesis were from the beginning rather 
controversial, and 
those who ``believed'' in quarks as real objects were sometimes
called ``frogs sitting 
around a pond saying  {\it quark, quark,quark} ``.

One important point is that we have baryons with spin 3/2,
like the  four $\Delta$-particles, having also isospin 3/2, meaning that there 
are four $\Delta$'s, with charges plus two, plus one, zero and minus one.
Assuming the $SU(3)$ triplet in Fig 6 build up all the baryons, there are three
cases where baryons are built up from three $u$-quarks, three $d$-quarks
 or three $s$-quarks, all with spin plus 1/2 {\it along the same axis.}
BUT: Three identical particles in the same state should be forbidden 
according to the {\it   Pauli exclusion principle;- ....unless..... there is
 another quantum number which is different for the three particles.}
This new quantum number introduced to solve the problem is for some
peculiar reason called {\it color}.
 And in order
to save the Pauli principle, there must be at least three different values 
of this new quantum number. Further,  if we say that there are just three
 different colors, it is natural to choose $SU(3)$ as the color symmetry 
group. In this case we write $SU(3)_c$. Then the
 matematicians can tell us that by combining three color triplet quarks
 we may obtain a {\it color singlet} (color neutral) baryon state.
 With other words, 
the color singlet has {\it no open color !}.

\subsection{ An elektron-proton collision. Partons }

The important question was: 
Is the proton elementary, or does it consist of smaller pieces?
To try to find out this there was
from 1969  going on an experiment at SLAC(California) were energetic electrons
were shot against protons or neutrons:
 
\begin{figure}
\begin{center}
\scalebox{0.45}{\includegraphics{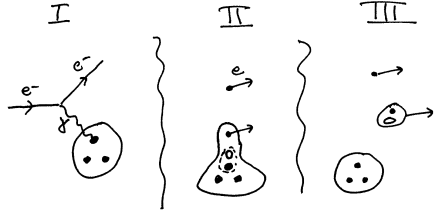}}
\caption{Elektron-proton collision at high energy 
at Stanford Linear Accellerator Center, in California, in 1969.
The process is $e \, p \rightarrow e \, N \, \pi$. Stage I: The incoming
electron hits-via a virtual photon-a quark inside the proton. Stage II:
Some of the energy transfer from the collision is used to make a
quark-antiquark pair in the proton. Stage III: The antiquark combine
with one of the quarks to make a meson which leaves the proton(more
general, the nucleon).
  } 
\label{epKoll}
\end{center}
\end{figure}

What was seen was that extra  particles could come out of the collision,
 for instance that 
 an extra pion was coming out, i. e. the inelastic
 reaction $e p \rightarrow e p \pi^0$
or $e p \rightarrow e n \pi^+$.
But the energetic  electron could never  shoot loose a  quark !.
The explanation 
was that the energy from the (virtual) electron may create a
quark-antiquark pair ($(q-\bar{q})$-pair) inside the proton. This
$(q-\bar{q})$-pair may come out as a {\it meson}
which is colorless.
A quark cannot be free! The  ``color''-charge is confined !
Quarks were always bound to a baryon or a meson.

\begin{figure}
\begin{center}
\scalebox{0.3}{\includegraphics{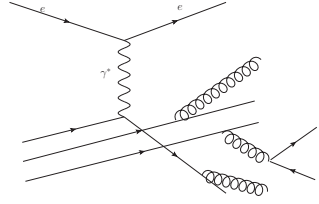}}
\caption{Collision $e p \rightarrow eX$ in the  Parton Model
  }
\label{epXdiagram}
\end{center} 
\end{figure}
It was also  seen that 
the scattering amplitude  contained, in addition to well known
kinematical factors,
  some  {\it structure functions} $f_p(x)$,
 which  depended 
only on the momentum $fraction$
$x \equiv p_{part}/p_{proton} = (-q^2)/(2 M_p \nu)$, 
of the partons (quarks). Here
$\nu = E_e - {E_e}^{'}$ is the energy difference between the in- and
out going electrons.

The intepretation of data is the following: At high energies, the proton
 consist of ``sea-quarks'' = {\it quarks} and  {\it anti-quarks} and 
 {\it gluons} in addition to the three {\it valence quarks} $(uud)$ forming
 the proton.
 In other words,  the proton contains the (usual) three valence quaks for
 small energies 
plus some gluons plus some  quark anti-quark pairs when the energy grows.
 The quarks cannot be kicked out of the proton. The energy
from the colliding electron (transmitted via the virtual photon) is used to
 make a quark anti-quark pair. Then for a (very) short time the proton
 contains four quarks and an anti-quark, which combine to one  baryon  and one 
meson. The quark anti-quark pair is thought to be made via the emission of a 
{\it gluon} which again makes the quark anti-quark pair. This gluon is the
 analogue of  the photon in elecromagnetic interactions, where a photon 
can make an electron-positron pair.

In 1974 came the  ``1974-revolution'':  Then the experimentalists found 
the socalled $\Psi$-particles. The decay-spectrum of these particles behaved 
qualitatively like the 
decay-spectrum  of
positronium $e^+ \, e^-$. The $\Psi$ particles were intepreted as 
a system of a heavy (for that time)  particle (quark) $c$ and its
anti-particle $ \bar{c}$. In other words $\Psi = (c \bar{c})$.
After these discoveries, the quark picture was accepted among (almost) all
 particle physicists.

\subsection{Nuclear forces}

Before the quark picture was accepted, strong interactions  meant  
nuclear forces. Now we might look slightly different about this.

 Two protons will repel each other electrically. But if they 
come very close the may attract each other because the color forces 
are stronger than the electric ones at very close distance.
Two protons will feel  the ``tail'' of quark-gluon-forces, in spite of
  electric repulsion.  \\
Nuclear forces are much stronger than  electromagnetic forces\,-
{\it -when the nucleons are close to each other}.

The color forces  are so  strong, for energies less than say 1 GeV,
that the quarks can not come out  of the proton.\\
The proton (-and the neutron) are color neutral.
Now the picture is that the true {\it strong force is the color force},
while the nuclear force is only the tail of the color force, some times 
called the {\it deduced color force}.
The color force binds quarks together to baryons and mesons, while
 the nuclear force bind nucleons to various clusters of nucleons, i.e. nuclei.

 What is strange about the color force is that it is relatively weak for
 high energies (corresponding to short distances) while it is extremely
 strong at long distanes
 (energies of about say hundred of MeV and smaller).
 When quarks were established
 there was 
a good candidate for a field theory for the color force,
namely the {\it Yang Mills theory}.

\section{Yang Mills Theory}

The Yang-Mills (YM) theory, which is used in modern field theory within 
the Standard model,  was introduced in 1954 already \cite{Yang:1954ek}.
The YM field theory came in some sense ``too early''.
It was apparently no use for it at the time it was presented in 1954.
There were no known massless vector bosons, 
except the photon. For instance, the $\rho$ had a mass, but the
YM theory has no mass term 
for the vectors. That would break gauge invariance.
 Was the YM theory  just a mathematical exercise?

 When I was a student in the sixties I  heard almost nothing
 about it. But knowledge
about it became extremely important in the mid seventies and later.
Having accepted the quarks, it was
clear that some variant of the YM theory might be used for interactions
between quarks mediated by gluons. It was potentially also   useful for weak
interactions,- if one found an acceptable way to handle massive vector
particles/fields. 
YM theories was a generalisation of the U(1) QED gauge theory to more
complex groups.

First, let us see what the Yang Mills theory is.
 YM theories have the   generic form:
\begin{equation}
{\cal L}_{YM} \, = \, \overline{f} \, \gamma^\mu \, 
\left(i D_\mu \, - m_f \right) f  - \frac{1}{4} F^{a, \mu \nu} \,  F^a_{\mu \nu}
\; \, ,
\end{equation}
where $f$ represents a multiplet of $n$ fermion (Dirac) fields, and
$D_\mu$ is the covariant derivative
\begin{equation}
i D_\mu \equiv i \partial_\mu \, -\,  g  \, A_\mu \; ,
\label{CovDer}
\end{equation}
where $g$ is the gauge coupling. This looks almost like QED,
but there is some important extra
ingredients.  Now the vector field $A_\mu$ is an $n \times n$ matrix
\begin{equation}
A_\mu = t^a \, A_\mu^a
\label{Afield}
\end{equation}
containing $N$ vector fields, and $t^a$ are the $N$
generators of the gauge group. If the gauge group is $SU(n)$, then
$N=(n^2-1)$.
Here the $t^a$'s are the generators of some gauge group.
Because the $t^a$ matrices do not commute among themselves, 
the field tensor  $F^a_{\mu \nu}$ has an extra term - to obtain gauge 
invariance, and is given by the commutator of two covariant derivatives: 
\begin{equation}
  [i \, D_\mu, i \, D_\nu] = - i g t^a F^a_{\mu \nu} \, , \; \,  
  F^a_{\mu \nu} \, = \, (\partial_\mu A^a_\nu \, - \, \partial_\nu A^a_\mu) \,
+ \, g \, f^{abc} A^b_\nu A^c_\nu
\label{YMFieldTens}
\end{equation}
It is important that the field tensor, unlike QED, has a quadratic term.
Squaring the field tensor to obtain the Lagrangian term for the vectors,
one  obtains triple
 an quartic terms. This makes YM theory qualitatively different from QED.
 Vector bosons will interact among themselves!  See Fig. 9 later.

\subsection{ Quantum chromo dynamics (QCD)}

Assuming that quarks exsist and carry three different colors,
one may write down  a quark gluon field theory of  Yang-Mills theory type
with gauge group $SU(3)_c$, where the subscript $c$ stands for {\it color}.
Thus QCD  has the form
\begin{eqnarray}
 {\cal L}_{QCD} \, = \, \sum_q \overline{q} \, \gamma^\mu \, 
\left(i \partial_\mu \, - \, g_s t^a \, A_\mu^a - m_q \right) q \, - \, 
\frac{1}{4} F^{a, \mu \nu} \,  F^a_{\mu \nu}
\label{QCDLag}
\end{eqnarray}
Note that there is a 
sum over quark flavors $q=u,d,s,...$, and that the gluons are the same 
for all flavors !.
The field tensor has an extra term, as all YM theories, see
 (\ref{YMFieldTens}):
\begin{equation}  
  F^a_{\mu \nu} \, = \, (\partial_\mu A^a_\nu \, - \, \partial_\nu A^a_\mu) \,
+ \, g_s \, f^{abc} A^b_\nu A^c_\nu \; ,
\label{GlueTens}
\end{equation}
where $A^a_\mu$ are the eight gluon four vector fields ($a=1,2,..8$ and
$\mu =0,1,2,3$).
The symbol $q$ is now very compact.
All the six quark fields $q$ ($q=u,d,s,c,b,t)$ have 
$3$ different color components (one Dirac field per color, and all Dirac
fields have four components). Thus the  quark field $q$ has in total
$6 \times 3 \times 4$ components.
The $t^a$ matrices are generators ($3 \times 3$-matrices) of the 
gauge group  $SU(3)_c$.
It is important that  the gluons interact among themselves! We obtain triple
 and quartic vertices. The existence of this triple gluon vertex has dramatic
 consequences for QCD compared to QED.

\begin{center}
\begin{figure}[htbp]
 \scalebox{0.4}{\includegraphics{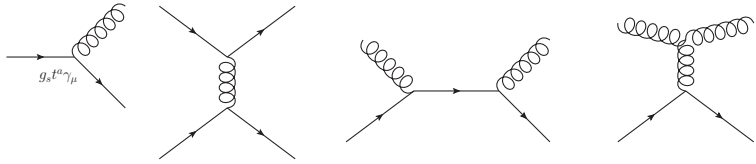}}
\caption{The quark gluon coupling(left diagram),Quark-quark scattering 
  (second diagram), and quark-gluon scattering (third and fourth diagram).
  The first three diagrams have an analog in QED, while this is not the case
  with the fourth diagram including the triple gluon coupling.}
\end{figure}
 \end{center}

QCD is significantly different from QED due to the triplet gluon coupling
The  copling $g_s$ is {\it universal}, the same for quark-gluon
and triple gluon couplings. This is a consequence og gauge invariance.
It is found that 
perturbative QCD breaks down for low energies (say 1-2 GeV. See later).

\subsection{Gauge transformations}

Both in QED and Yang Mills theory (-and thereby QCD) an important
ingredience is gauge invariance.

The gauge transformations in QED are
 \begin{eqnarray}
 \psi(x) \; \rightarrow  \; \psi(x)'  \, =  \, e^{i \alpha(x)}
\,   \, \psi(x) \; , \, A_\mu(x)  \rightarrow
 A_\mu(x)' \, = \,  A_\mu(x) \, - \, \frac{1}{e} \, \partial_\mu \alpha(x) \; .
\label{gaugeQED}
\end{eqnarray}
 Under these combined transformations the QED Lagrangian
 given by (\ref{LQED}) - (\ref{IntTerm}) is invariant.
 
 For Yang-Mills theory the corresponding transformation for fermion fields is
 \begin{eqnarray}
 f(x) \; \rightarrow  \; f(x)'  \, =  \; U(x) \, f(x)
 \quad ; \; \, U(x) \, \in \, SU(n) \;  (-\mbox {for instance})  \; .
\label{qtr}
\end{eqnarray}
 Gauge field transformations  in the non-Abelian has to be such that
 $D_\mu f$ tansforms as $f$:
 \begin{equation}
 i D_\mu f \; \rightarrow i D'_\mu f' \, = U \,  i D_\mu f \; .
\label{Dqtrans}
\end{equation}
To obtain this the transformation must  contain a 
rotation among fields: 
\begin{equation}
 A_\mu  \,  \, \rightarrow A_\mu' \;
= \; U \, A_\mu \, U^\dagger \; + \; \frac{i}{g} \,
\left(\partial_\mu U\right) \, U^\dagger,   
\label{Atrans}
\end{equation}
and 
\begin{equation}
   F_{\mu \nu} \,  \rightarrow \, F_{\mu \nu}' \; = \; 
   U \, F_{\mu \nu} \, U^\dagger \; \,  ;  \mbox{where} \; \,
   U \, = \, U(x) \, = \, 
e^{i t^a \alpha^a(x)} \; ,
\label{tensorTrans}
\end{equation}
where $\alpha^a(x)$ is a  set of real  functions. 
 Here $A(x)_\mu   \equiv  t^a A(x)_\mu^a \,$, and similarly we define
 $F(x)_{\mu \nu}   \equiv  t^a F(x)_{\mu \nu}^a$.

\section{Comparison of running couplings in QED and QCD}

\begin{center}
\begin{figure}[htbp]
 \scalebox{0.35}{\includegraphics{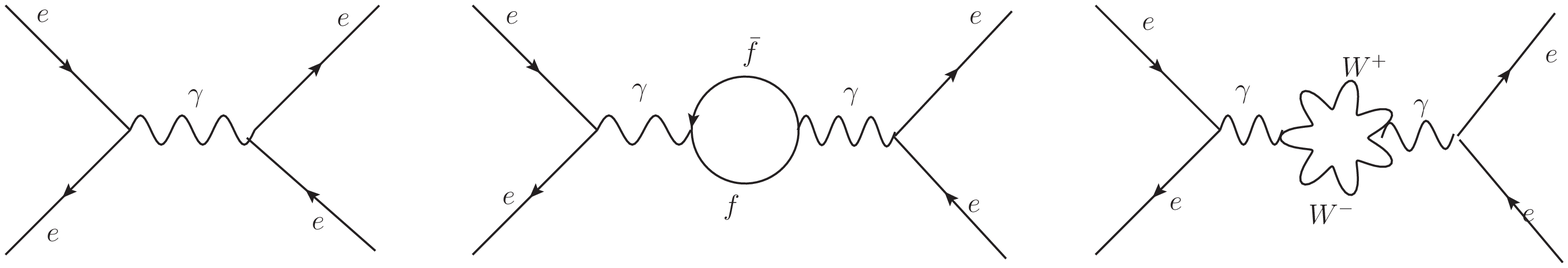}}
\caption{Diagrams explaining $\alpha_{em}$, which grows with energy!
  $\alpha_{em} \simeq 1/129$ at energy = $M_Z c^2$.
  Alternatively, in earlier times this was considered as a
  modification of Coulomb potential}
\label{EWcorr}
\end{figure}
 \end{center}

\begin{center}
\begin{figure}[htbp]
 \scalebox{0.35}{\includegraphics{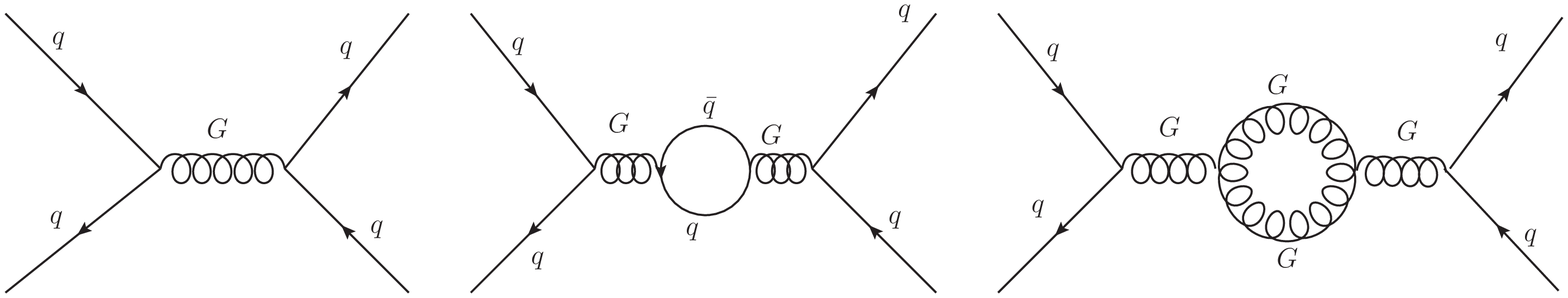}}
 \caption{Diagrams explaining the variation (``running'') of
   $\alpha_{s}$, which become smaller with
 growing energy. The quark and the gluon loops have different signs. 
The gluon loop dominates. This has dramatic consequenses for the high-energy 
behavious of perturbative QCD. The last diagram dominates over the diagram
in the middle.}
\label{qq-corr}
\end{figure}
 \end{center}

Quantum fluctuations means that particles suddenly appear and  
might disappear again. Such effects are visualized by loop effects.
In Fig. \ref{EWcorr} we display $e e$ scattering to lowest order (left)
then the possible fermion loop corrections
($f$ means a charged fermion), and to 
the right a diagram with a $W W $ loop, because $W$s are electrically charged.
In Fig. \ref{qq-corr} we show $qq$-scattering to lowest order in 
perturbative QCD (left) with corrections with a quark loop(middle) and 
gluonic loop. (right).

 \begin{center}
\begin{figure}[htbp]
 \scalebox{0.4}{\includegraphics{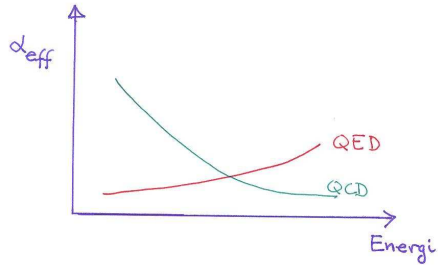}}
\caption{Qualitative behaviour  of effective
 $\alpha_{em}$ and of effective $\alpha_s$ with variation of energy}
\end{figure}
 \end{center}

 It is found that perturbative QCD  breaks down at small energies! Then other
 methods are needed.
 Lattice gauge theory is considered to be the best method. Earlier some
 variant of quark models also were used. 

 Experimentally one found:
\begin{equation}
\alpha_s(M_Z^2) \simeq 0.12 \; ,
\label{AlphaZ}
\end{equation}
Using this as a starting point, one could use the calculations of
the diagrams in Fig. \ref{qq-corr} to find:   
  \begin{equation}
    \alpha_s((1GeV)^2) \simeq 0.5 \, (\pm 10 \%) , \; \, \mbox{and} \; \,
 \alpha_s((0.7GeV)^2) \simeq 0.6 \; \mbox{to } 1.2
\label{alphas}
  \end{equation}
 which illustrates the breakdown of perturbative QCD at energies
 below, say, one GeV.

\section{$SU(3)_L \times SU(3)_R$
 - symmetry and chiral perturbation theory}

For 
QCD with the  three lightest flavors $q=(u,d,s)$ (with quark
masses below one GeV)
in the Lagrangian might be split in left and right-handed fields: 
\bea
{\cal L}_{QCD} \; = \; 
\overline{q_L} \left(\gamma \cdot i D \right) \, q_L \, + \, 
\overline{q_R} \left( \gamma \cdot i D \right) \, q_R \, + \, 
 m_q ( \overline{q_L} \, q_R \, + \, \overline{q_R} \, q_L )
 + \, {\cal L}_{G} \; ,
 \eea
 where $\, {\cal L}_{G} $ is the gauge (gluon) part.
Here $q= q_L + q_R$ .

Without the mass term, (i.e for  $m_q \rightarrow 0$)  
this Lagrangian is invariant under the global transformations
\bea
q_L \, \rightarrow  \, V_L \, q_L \quad ; \;  
q_R \, \rightarrow  \, V_R \, q_R \quad ; \; V^\dagger_L \, V_L \, = \,  
 V^\dagger_R \, V_R \, = 1 \; ;  \; 
V^\dagger_L V_R \neq 1 \, .
\label{GlobalSU32}
\eea
Thus there is  a chiral $SU(3)_L \times SU(3)_R$ symmetry.
Then one might have expected parity doublets, say a proton-like state
with opposite parity. Such states does not exists. So in some way this
symmetry must be broken.
The solution is that we have {\it spontanous symmetry breaking}
(SSB) in QCD . This appear as a 
quark condensate:
\bea
\langle \bar{q} q \rangle \, \equiv \, 
\langle \, 0| \, \bar{q} q \, | \, 0  \rangle \; 
\simeq (-240 \, \text{MeV})^3
\neq \, 0 \, .
\label{qCondensate}
\eea
QCD has a non-trivial vacuum!
The  $SU_L(3) \times SU_R(3)$ symmetry of QCD (for $m_q
\rightarrow \, 0$)  {\it breaks down}.
There is also a gluon condensate
\bea
\langle \frac{\alpha_s}{\pi}G^2 \,\rangle 
 \sim (300 \; \text{to} \; 400 \; \text{MeV})^4 \neq \; 0  \; .
\eea
The  Goldstone bosons of this spontanous broken symmetry is the
pseudoscalar meson octet ($\pi,K,\eta_8$).

Now one may write down the chiral perturbation theory for the
Goldstone particles:
\begin{equation}
{\cal L}_{\chi PT} \, = \, 
\frac{f_\pi^2}{4} Tr[(\partial_\mu U) (\partial^\mu U^\dagger)] + .....
\label{chiPT}
\end{equation} 
where $f_\pi$ is the decay constant from  
the decay $\pi \rightarrow \mu \nu_\mu$  and 
\begin{equation}
U \, \equiv \, exp{(i \lambda^a \, \pi^a/f_\pi)} \; ,
\end{equation}
where $\lambda^a$ are generators for the flavor group $SU(3)_F$, and
the sum runs over the octet ($\pi,K,\eta$).
One has:
\begin{equation}
 m_{\pi}^2 \, \sim \, - \, \langle \bar{q} q \rangle \
\, m_{u,d} / f_\pi^2,
\end{equation}
This is something new. In pure perturbative theories, physical quantities
always
depend on the square of quark masses.
We obtain a
$SU(3)_F$-symmetric chiral perturbation theory ($\chi PT$) for mesons and 
baryons which contains a lot of terms. This  theory is
{\it non-renormalizable}.\\
But still $ \chi PT$ is useful, for instance in decays of $K$-mesons.
Various quark models, say the chiral quark model ($\chi QM$) can be
 used to give an  estimate of some coefficients
of terms.($\chi QM$ connects quarks to mesons)

\section{Weak interactions}

When I was a young student in the 1960s I learned little about weak
interactions.
Maybe it was just because it was  {\it weak}- and not so much to worry about
compared to electromagnetic and strong interactions.
This changed in the mid 1970s when it was growing more and more clear that
weak interactions might be described by field theoretic models.

But let us start with the beginning.
Before we continue, we have to remind ourselves that 
before 1956, the discrete symmetries {\it C-, P-, T}
was thought to hold in all interactions. More concrete,
so far this was thought to hold separately in 
 gravity, electromagnetic and strong interactions
 as well as weak interactions.

\begin{itemize}
\item 
First, a reminder: \\
$C$ = charge conjugation take  a particle to an anti-particle\\
$P$ = Parity(mirror) transformation  reverse the direction of a vector: 
 $\vec{r} \rightarrow  \,- \, \vec{r}$\\
(where $\vec{r}$ is the  position vector)) \\
For an axial  vector like angular momentum: $\vec{L} = \vec{r} \times \vec{p}$
$\rightarrow + \vec{L} $ does not change sign under a parity transformation\\
$T$ = Time reversal reverses time: $t  \, \rightarrow  \, - \, t \, $ 
(Remember : In quantum mechanics the $T$-operator is anti-unitarian and 
changes sign of the imaginary unit $i$)

\end{itemize}

\subsection{Parity symmetry is broken in weak interactions(decays)}

Early in the 1950's before I enteed university physics,
one discussed the 
``$\theta-\tau$-puzzle'':
The experimentalists  found apparently two different particles
 $\theta$ and $\tau$
 with the same mass.  The $\theta$ decayed to two pions ($\pi$-mesons), and  
$\tau$ deayed to 3 pions. $\theta$ and $\tau$ was thought to be different
 because a system of two pions has different parity compared to a system
 of three pions. 
Now Lee and Yang  explained in 1956 that the solution to the
 puzzle is that
 parity is a {\it  broken symmetry in weak interactions}
 and that $\theta$ and $\tau$
is actually the same particle, namely the {\it kaon} ($K$-meson), which 
could decay both to two or three pions.
The  $\pi$'s  and $K$'s are pseudo-scalar particles, that is, thay have 
spin zero and intrinsic parity minus ($0^-$).

\subsection{Early Weak decays}

The most known weak process  is probably $\beta$-decay : 
\begin{equation}
n \rightarrow p \, e^- \; \overline{\nu_e} \; ,
\label{betadecay}
\end{equation}
which may occurr for free neutrons or for neutrons bound in a nucleus.
The process is famous because of the $C-14$ age test.
A free neutron has an average life time of order  15 minutes. For a bound
neutron it depends on the nucleus. In many cases it is stable,
and is not a $\beta$-emitter. In a few nuclei the  bound proton might
decay (they are positron emitters) and we have the inverse process:
\begin{equation}
p \rightarrow n \, e^+ \nu_e  \; .
\label{pbetadecay}
\end{equation}
This cannot occurr for free protons because the proton is
lighther than the neutron. The process (\ref{pbetadecay})
is used in PET-scanners in hospitals.
The process(\ref{betadecay}) and (\ref{pbetadecay})
 also breaks parity-symmetry.

The charged $\pi$'s and $K$'s could also decay to leptons, for instance 
\begin{equation}
\pi^+ , K^+ \rightarrow \mu^+ \nu_\mu  \; \, \mbox{or} \; \, e^+ \nu_e
\; .
\label{piKdecay}
\end{equation}
Here the decays to the heavier lepton $\mu$ is most probable.
As a curiocity: When the muon was discovered one physicist said:
``Who ordered that''. He meant it was ``no need'' for it.
However, it is useful because then pions from cosmic radiation decay
faster than if the electron were the only charged lepton. With only
the electron we would get more pions in our heads!

For massive fermions and anti-fermions there are in total four degrees
of freedom.
BUT: For massless fermions/anti-fermions there are in total {\it only  two
 degrees of freedom }
Then one makes the following definitions: 

Neutrinos are considered  to be lefthanded particles $(\nu_e)_L$,
and and antineutrinos as righthanded antiparticles 
$(\overline{\nu_e})_R$. Then one defined the leptoic charges for the
electron, the muon ($\mu$) and their neutrinos.Then one has:  
 
\begin{itemize}
\item
Lepton number $L_e$ is conserved: \\
$L(e^-) \, = \, L(\nu_e) = \, + 1$ and
 $L(e^+) \, = \, L(\overline{\nu_e}) = \, - 1$
\item
Lepton number $L_\mu$ conserved: \\
$L(\mu^-) \, = \, L(\nu_\mu) = \, + 1$ and
 $L(\mu^+) \, = \, L(\overline{\nu_\mu}) = \, - 1$\\
(Similar for $L_\tau$) 
\end{itemize}
such that $L_e$ and $L_\mu$ was (so far) conserved.
Thus one had
conservation of lepton numbers  built into the theory !
-as visible in the pure leptonic decay mode
\begin{equation}
\mu^- \rightarrow \, e^- \overline{\nu_e} \, \nu_\mu \; .
\label{muedecay}
\end{equation}

In the early 1960's there were no  theory for weak interactions like 
QED for the electromagnetic interactions.
But already in the 1930-ties
Fermi wrote down an effective interaction for four fermion fields
acting in a point.
Thought as an interaction Lagrangian, the Fermi interaction contained a
product of the neutron field $\psi_n$, the (adjoint of) the proton field
$\psi_p$ the electron field $\psi_e$ and the (adjoint of) the neutrino
field $\psi_\nu$.
But was the fields combined into vectors (as in QED), or was it scalars,
or tensors?
Little was from the beginning known about the interaction, except for its
strength,
the {\it Fermi constant}
\begin{equation}
G_F \simeq \frac{10^{-5}}{m_p^2} \; \, .
\end{equation}
Later, in 1958  Feynman and Gell-Mann showed \cite{Feynman:1958ty}
that the interaction could
be written as a product of two
{\it left-handed} currents
\begin{equation}
  {\cal L}_F \, = \, 4 \frac{G_F}{\sqrt{2}} \, j(W,N)^\mu \,
  j(W,l)_\mu \; , \quad
\label{FermiTh}
\end{equation}
where the weak currents are left-handed,- i.e a vector current minus 
an axial current divided by 2:
\begin{equation}
 j(W)_\mu \, = \, \frac{1}{2} \left( j^V_\mu \, - \,  j^A_\mu \right) \, .
\label{FermiTh2}
\end{equation}
 In beta-decay the nucleonic (N) and an leptonic (l)  currents can be written
\begin{equation}
j(W,N)^\mu = \frac{1}{2}\overline{\psi_p}\gamma^\mu(1-\gamma_5)\psi_n \; \; 
\; \, , \; \, \mbox{and} \; \, 
j(W,l)^\mu = \frac{1}{2}\overline{\psi_e}\gamma^\mu(1-\gamma_5)\psi_{\nu_e} \,
\; ,
\label{Fermicur}
\end{equation}
respectively. The $\psi$'s are the various particle fields in a 
self-explained way.
Note that $ j^A_\mu \, = \, \overline{\psi} \, \gamma_\mu \,
\gamma_5 \, \psi$, such that
the left-handed current may be written:
\begin{equation}
j(W)_\mu \, = \psi \gamma_\mu \, P_L \, \psi \; , \quad 
 P_{L,R} \, \equiv \,  \frac{1}{2} \, \left( 1 \, \mp \, \gamma_5 \, \right) 
\label{Left-cur}
\end{equation}
where $P_{L,R}$ are the left and right projectors in Dirac space:
\begin{equation}
 (P_{L})^2 \, = \,  P_{L} \; , \; \,  (P_{R})^2 \, = \,  P_{R} \; , \;
 P_{L} \cdot  P_{R} \, = \, P_{R} \, \cdot P_{L} \, = 0 \, , 
P_{L} \, + \, P_{R} \, = \, 1 \; .
\label{Proj}
\end{equation}
A Dirac field can be split in two parts:  
\begin{equation}
\psi \, = \, \psi_L \, + \psi_R \; , \; \psi_L \, = \, P_L \, \psi \; , \; 
\psi_R \, = \, P_R \, \psi \; ,
\label{Proj2}
\end{equation}
where $\psi_L$ = left spinning  (left screw)particle.
and  $\psi_R$ = right spinning (right screw) particle.
In this way parity violation is built into the theory.
Note  that if the neutrino $\nu_e$ is purely lefthanded, then
the apparent vector  current
for $\nu_e \rightarrow e^-$ is automatically lefthanded:
\begin{equation}
j(W,e)_\mu \, = \, \overline{\psi_e} \, \gamma_\mu \, \psi_{\nu_e} 
\; = \, \overline{\psi_e} \, \gamma_\mu \, (\psi_{\nu_e})_L
\; = \, \overline{\psi_e} \, \gamma_\mu \, P_L \psi_{\nu_e} \; .
\label{curnu}
\end{equation}
Note that $\overline{\psi_L} = 
(P_L \psi)^\dagger \gamma_0 = \overline{\psi} \, P_R$, such that:
\begin{equation}
\overline{\psi_L} \, \gamma_\mu \, \psi_L  \, = \, 
\overline{\psi} \, \gamma_\mu \, P_L \psi \; \, , \; \,
\overline{\psi_R} \, \gamma_\mu \, \psi_R  \, = \, 
\overline{\psi} \, \gamma_\mu \, P_R \psi \; \, , \; \;
\overline{\psi} \, \psi \, = \, \overline{\psi_L} \,  \psi_R \, 
\, + \, \overline{\psi_R} \,  \psi_L \; .
\label{RLcur}
\end{equation}
From the last equation we observe that mass terms mixes left and right fields.
We observe that symbolically for a product of lefthanded currents:
\begin{equation}
(V-A)^2 \; = \; (V\, V \, + A \, A) \; - \; (V \, A \, + A \, V)  \; ,
\label{VA2}
\end{equation}
where the last term violates parity !

\begin{figure}
\begin{center}
\scalebox{0.4}{\includegraphics{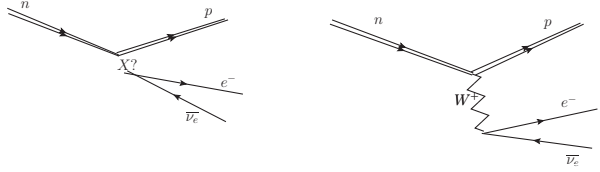}}
\caption{Diagram for  beta-decay in Fermi-theory(left).
  Hypothetical extension with a $W$-boson (right).
  }
\label{FermiDiagram}
\end{center}
\end{figure}
If the two currents were mediated by a heavy weak boson $W$ coupling to 
 with strength $g_W$, then one would have  
\begin{equation}
\, - \, 4 \frac{G_F}{\sqrt{2}} \; \, \rightarrow \; \, 
\frac{g_W^2}{q^2- M_W^2}  \, .
\label{Fermi-W2}
\end{equation}
If the four momentum $q$ is small compared to $M_W$; i.e $|q^2| << M_W^2$, then
\begin{equation}
  4 \frac{G_F}{\sqrt{2}} \, = \,  \frac{g_W^2}{M_W^2} \; .
  \label{Fermi-W2b}
\end{equation}
There existed an hypothes already in the 1930's  that weak
interactions were mediated by a massive boson $W$.  
But at that time the $W$ was very hypothetic, and how massive it should be
was not known. (its spin was also not clear). But if one assumed that $g_W$ were of the same order of
 magnitude as the electric charge $e$, one could deduce from the measured 
 value of $G_F$, that the mass of the $W$ would be of the order
 40 proton masses, which turned up to be correct up to a factor 2.

\subsection{CP-symmetry believed to be true !}

As parity($P$)- and charge conjugation ($C$)-symmetry was  found to be
broken, it was proposed that the combined $C P$-symmetry could be a
 true symmetry. For instance the russian physicist Lev Landau argued for 
this solution. To understand that CP-symmetry  is a useful concept, one may 
look at leptonic decays of pions in the following illustrative example:
\\

\begin{figure}
\begin{center}
\scalebox{0.6}{\includegraphics{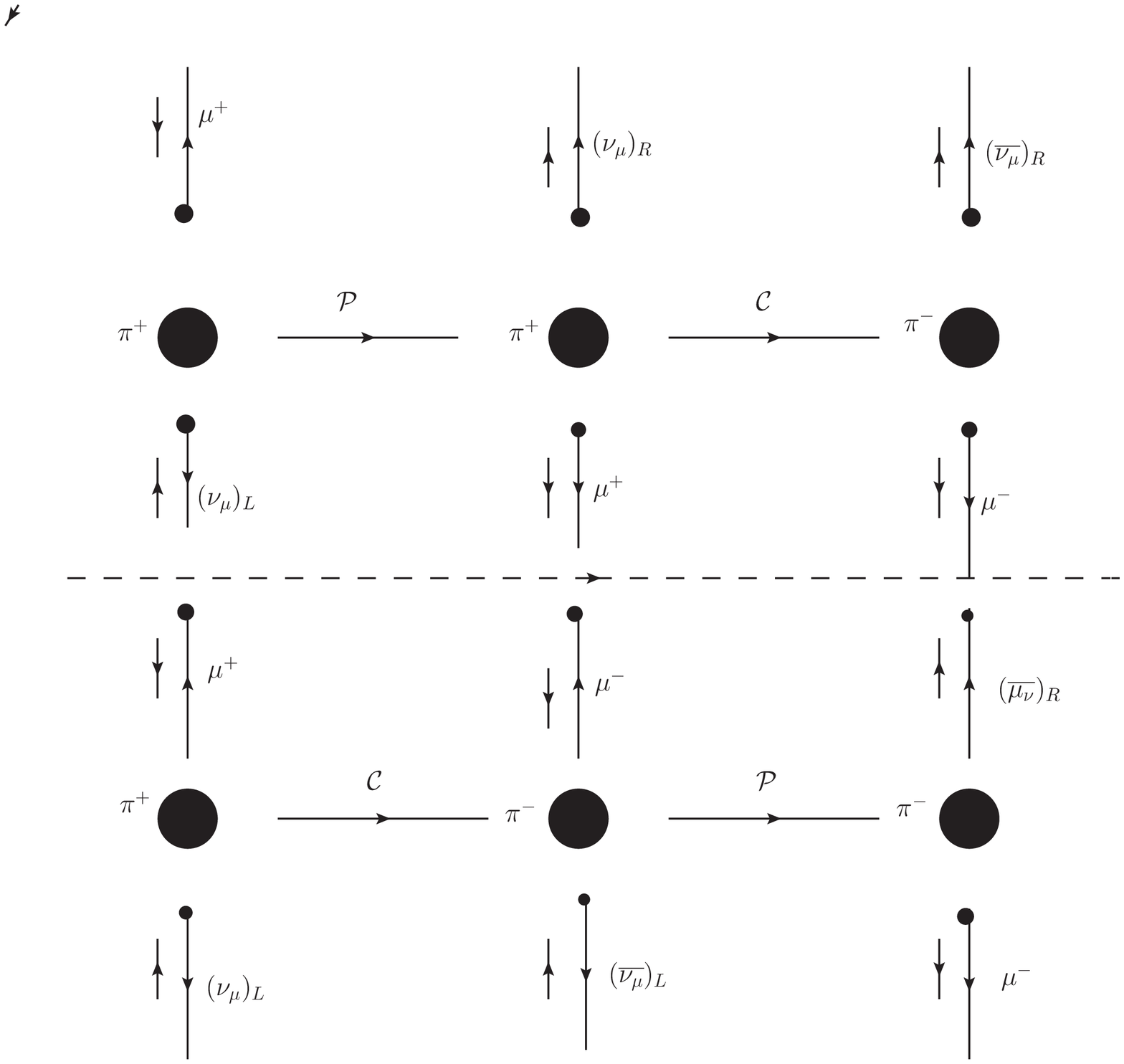}}
\caption{$CP$-transformation of $\pi^+ \rightarrow \mu^+ \, \nu_\mu$.
  Above the horisontal dashed line the parity transformation ${\cal P}$
  is performed before charge conjugation ${\cal C}$. Below the horisontal
  dashed line, the charge conjugation is performed before the parity
  transformation. The  big blobs are
  pions, and the smaller leptons. The long vertical lines(arrows) are
  lepton momenta, and the short arrows are lepton spins.
  }
\label{CP-transform}
\end{center}
\end{figure}

The simplest decays of pions are $\pi^+ \rightarrow \mu^+ \, \nu_\mu$ and 
$\pi^- \rightarrow \mu^- \, \overline{\nu_\mu}$. 
If $CP$ symmetry is valid, these processes should have the same probabilities.
namely, let us look at $C$ and $P$-transformations separately:
As said above, the neutrino is left-handed, meaning that the spin of 
the neutrino is opposite of the momentum. 
First we consider a parity transformation of the process
 $\left[\pi^+ \rightarrow \mu^+ \, (\nu_\mu)_L \right] $. Because the pion
 has zero spin,  the spin of the $\mu^+$ is directed opposite to the
 neutrino-spin. This means that here the  $\mu^+$ is a left-handed particle
$(\mu^+)_L$.  A parity transformation will reverse
 the momenta but not the spins,
 leading to the ``process''
 $\left[ \pi^+ \rightarrow (\mu^+)_R \, (\nu_\mu)_R \right]$.
 But a right-handed neutrino is here unphysical.
 And this process 
is not possible (in the limit of zero neutrino mass).
{\it However}, adding a charge conjugation, particles go to antiparticles.
Then we obtain a right- handed anti-neutrino, which is  physical, such that
the $CP$ transformated process is
$ CP [\pi^+ \rightarrow (\mu^+)_L \, (\nu_\mu)_L ] \; $ = 
$C [ \pi^+ \rightarrow (\mu^+)_R \, (\nu_\mu)_R ] $  
 = $ \, [\pi^- \rightarrow (\mu^-)_R \, (\overline{\nu_\mu})_R]$. 
One may also make the transformations in opposite order:
The C-tansformation is 
 $C\left[\pi^+ \rightarrow (\mu^+)_L \, (\nu_\mu)_L\right]$ = 
 $ [\pi^- \rightarrow (\mu^+)_L \, (\overline{\nu_\mu})_L +]$.
But here a left-handed anti-neutrino does not exist, so this process 
does not exist. Adding now the parity-transformation, we obtain
further
$P\left[\pi^- \rightarrow  (\mu^+)_L \, (\overline{\nu_\mu})_L \right]$
$ =  \, \pi^- \rightarrow (\mu^-)_R \, (\overline{\nu_\mu})_R$,
 which is a physical process with the same probabillity as the original 
The conclusion is that the processes
$\pi^+ \rightarrow \mu^+ \, (\nu_\mu)_L $ and  
$\pi^- \rightarrow \mu^- \, (\overline{\nu_\mu})_R$ should have the same 
probability.

If so, we may say there is a symmetry between left-handed matter 
and right-handed anti-matter, {\it viz.}:
\begin{equation}
(matter)_L \; \, \leftrightarrow \; \, (anti-matter)_R \; \, .
\end{equation}

\subsection{CP-symmetry is also broken in weak interaction}

A second shock in particle physics community came in 1964 in the experiments
 by Fitch, Cronin and coworkers \cite{Christenson:1964fg}
 which showed that also $CP$-symmetry is broken. As I entered university
 physics in 1963, I did not register this shock from the beginning.
Because the electrically neutral $K$-mesons $K^0$ and $\overline{K^0}$ are 
antiparticles 
of each other, they  are degenerate in mass.
 Also because they are electrically neutral they may form mixtures 
which are physical  
$CP$ eigenstates (with a mixing angle near 45 degrees):
\begin{equation}
 |K_\pm> \, = \, (|K^0 > \, \mp \, |\overline{K^0} >)/\sqrt{2} \; \,
\mbox{ leading to} \; \;  CP|K_\pm> \, = \, \pm |K_\pm> \; \, .
\label{Kstates}
\end{equation}
As in ordinary quantum mechanics, if there is an interaction which takes 
$K^0$ into $\overline{K^0}$ and opposite, the degeneration is lifted.
 The 
physical states would be $K_S = K_+$ (shortlived) and $K_L = K_-$ (longlived).
 {\it If } $\, CP$-symmetry were fulfilled, then
we should see the decay modes  $K_S \rightarrow 2 \pi$ and
 $K_L \rightarrow 3 \pi$, 
because two pions  has positive $CP$-symmetry,  while three pions
may have negative 
$CP$-symmmetry.
But it turned out that this was not exact because  some 2 to 3 permille
 of $K_L$ decayed to $2 \pi$ !
The explanation for this is that $K_L$ will be a sum of $K_+$ and $K_+$
\begin{equation}
| K_L \rangle \simeq | K_-  \rangle \, + \, \epsilon \,  | K_+  \rangle \; \, 
\mbox{with} \; \, |\epsilon| \simeq 2.6 \times 10^{-3} \; \, .
\label{KLstate}
\end{equation}
Accordingly CP-symmetry is broken at a permille level.
Later we will look at the  explanation for this CP-violation mixing
parameter named $\epsilon$.

\section{Electroweak interactions}

In 1967, S. Weinberg published a paper with the title  ``A model of leptons''
\cite{Weinberg:1967tq} which was a model of weak interactions for leptons which was a
Yang Mills type {\it gauge theory} for 
the group theory of  $SU(2)_L \times U(1)_Y$ with two lepton families, the electron, the muon and their neutrinos. 
At roughly the same time  S. Glashow and A. Salam had similar ideas.

But this paper was not cited very much the first couple of years,
but after that
the citations exploded. It should be noted that at that time,
the $SU(2) \times U(1)$
version was not the only one. A version built on the group $O(3)$ and
including heavy lepton also existed.

Could one build an acceptable theory for weak interactions? Obviously the
$W$'s should be very heavy vector particles, unlike the photon
which is mass-less. Also the the $W$'s must be electrically charged
($W^\pm$), and
should the be expected to interact with the photon, as other charged particles.
Then there should not be a big surprice that one could build $W$'s
and the photon into the same theory, - in an {\it electroweak theory}.

\begin{figure}
\begin{center}
\scalebox{0.4}{\includegraphics{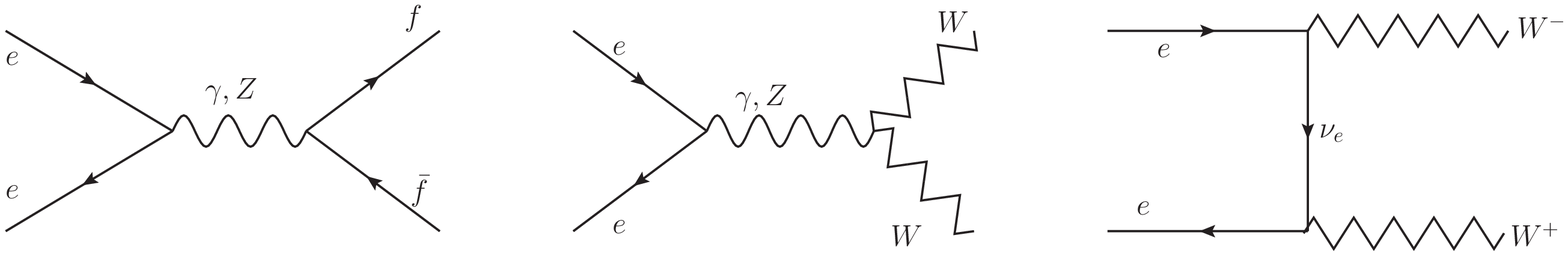}}
\caption{Examples: $ee \rightarrow f \bar{f}$ and $ee \rightarrow W^+ \, W^-$
  }
\label{eescatt}
\end{center}
\end{figure}
\begin{figure}
\begin{center}
\scalebox{0.4}{\includegraphics{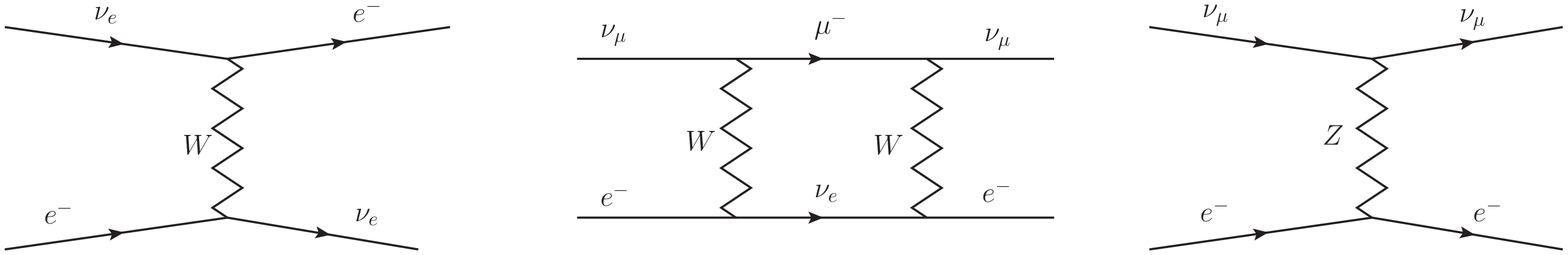}}
\caption{$\nu_e \, e^-$  scattering may occurr with the exchange of one
 $W$-boson (Left diagram) . If the $Z$-boson didnt exist, the similar
  $\nu_\mu \, e^-$ scattering had to proceed with exchange of two $W$-bosons
  (middle diagram). With the existence of the $Z$-boson, this latter
  process may proceed with
  the exchange of a single $Z$-boson (right diagram)}.
\label{numuescatt}
\end{center}
\end{figure}

A visual argument for this is to consider the process
$e^+ \, e^- \, \rightarrow \, W^+ \, W^-$ where both weak and
electromagnetic interactions are present, as shown in Fig \ref{eescatt}.
There were also
theoretical arguments that there should exist a neutral weak current
and that together with the charged $W$'s there should be a neutral
weak boson $Z$. Then a proces like
$\nu_\mu \, e^-\, \rightarrow \, \nu_\mu \, e^-$ can proceed with
exchange of a single $Z$-boson. Otherwise it had to proceed via exchange
of two $W$'s and would be much weaker. See Fig. \ref{numuescatt}.

\subsection{``The birth of a symmtry'' }

In Weinberg's model
the particle fields was split in right and left-handed particles.
Further, the left-handed particles were put in $SU2)_L$ doublets while the
right-handed components were $SU2)_L$ singlets.
The right-handed $e$ and $\mu$ fields are SU(2) singlets. 
Without right-handed neutrino fields, they are mass-less.
Mathematially, $W_\mu^i$ ($i=1,2,3)$ are the gauge bosons of $SU(2)_L$
and $B_\mu$ of $U(1)_Y$. Note that from now on the symbol $e$ means the
electron field, and the electric charge is called $g_\gamma$, which is the photon coupling to charged fermions.

 Here we consider one lepton doublet interacting with gauge bosons, \\
 namely:
 $E_L \, \equiv \, \left(\begin{array}{c}\nu_e\\ e^-\end{array} \right)_L$
 and a $SU(2)_L$ singlet $e_R$. 
There are four gauge bosons, the three  $W^i_\mu \; , i=1,2,3$ for $SU(2)_L$
 and $B_\mu$ for $U(1)_Y$.
The Lagrangian for one (the electronic) doublet is 
\begin{eqnarray}
 {\cal L}_{EW1l} \, = \, \overline{E_L} \, \gamma^\mu \,i D^{(L)}_\mu \, E_L
 \, + \, \overline{e_R} \gamma^\mu \, \left(i \partial_\mu \, -
 \, \frac{1}{2} \,g' \,  Y_R \, B_\mu \right) e_R  \nonumber \\
- \frac{1}{4} W^{i, \mu \nu} \,  W^i_{\mu \nu} \, - \, 
\frac{1}{4} B^{\mu \nu} \,  B_{\mu \nu} \, 
 \, + .......... \; ,
\label{EWLept}
\end{eqnarray}
where the covariant derivative
\begin{eqnarray}
  i D^{(L)}_\mu \, = \,
\left(i \partial_\mu \, - g \, \frac{1}{2} \tau^i W^i_\mu \, - \, 
\frac{1}{2} \,g' \,  Y_L \ B_\mu \right) 
\label{EWCovDer}
\end{eqnarray}
is $2\times2$ matrix (A $2\times2$ unit matrix in front of $\partial_\mu$ and
$Y_L$ is understood). Note that the $SU(2)_L$ coupling is $g$ and the $U(1)_Y$
coupling is $g'$.
The Lagrangian above may also be written as
\begin{eqnarray}
 {\cal L}_{EW1l} \, = \, \overline{E_L} \, \gamma^\mu \, 
 i \partial_\mu \, E_L \, + 
 \overline{e_R} \gamma^\mu \,  \partial_\mu \, e_R 
 \, - \, g \, W^i_\mu j^{i \mu} \, -
 \, \frac{g'}{2} \, B^\mu \, j^Y_\mu \nonumber \\
  \, - \frac{1}{4} W^{i, \mu \nu} \,  W^i_{\mu \nu} \, - \, 
 \frac{1}{4} B^{\mu \nu} \,  B_{\mu \nu} \,
 \, + ..... \; ,
.....
\label{EWLeptCur}
\end{eqnarray}
where
\begin{eqnarray}
j^i_\mu \, = \, \overline{E_L} \, \gamma_\mu  \,  
\frac{1}{2} \tau^i \, E_L \, \;  \mbox{and} \, \;
j^Y_\mu \, = \,Y_L \, \overline{E_L} \gamma_\mu \, E_L \, + \, 
Y_R \, \overline{e_R} \gamma_\mu \, e_R \; \, . 
\label{LeptCur}
\end{eqnarray}
are the weak isospin  and weak hyper-charge currents respectively.
$Y_L$ and $Y_R$ ae numbers (the values of the weak hypercharges)

All particles  are apparently massless in order to have 
 {\it gauge invariance !}.
The masses will enter through the {\it Higgs-mechanics} as we soon will show.

In the end,  the  weak hypercharges  must be chosen
such that we obtain the correct physical currents. It turns out
that we obtain this by the same formula as in (\ref{chargeForm})
\begin{equation}
Q \, = \, I_3^w \, + \frac{Y^w}{2} \; ,
\label{chargeFormW}
\end{equation}
where the superscript $w$ now stands for {\it weak}. Thus the physical
content is different than in equation (\ref{chargeForm}).

In the end the Lagrangian must be
\begin{eqnarray}
{\cal L}_{EW1l} \; = \; \overline{e} \, \gamma^\mu \,
 i \partial_\mu \, e \, + \, \overline{(\nu_e)_L} \, \gamma^\mu \,
 i \partial_\mu \, (\nu_e)_L  \,
- \, g_\gamma \, j^{em}_\mu  \, A^\mu  \, - \, g_Z \, j^{Z}_\mu  \, Z^\mu
\nonumber \\
 - \, g_W \, \left(j^{(+)}_\mu  \, W^{(-)\mu} \, + \,  j^{(-)}_\mu  \,
W^{(+)\mu}\right) \; + \, (W,Z,A)-terms + .... \; ,
\label{currtrans}
\end{eqnarray}
where $g_\gamma $ (= electric charge), $g_W$ and $g_Z$ are the couplings
of $\gamma$ (=photon), $W$, and $Z$ respectively.

 The  physical bosons  
$W_\mu^{(\pm)}, Z_\mu, A_\mu$ are given  in the following way:
\begin{equation}
W_\mu^{\pm} \; = \; \frac{1}{\sqrt{2}}(W_\mu^{1)} \pm i \, W_\mu^{2}) \; \, 
,  \; Z_\mu = c_W W_\mu^3 + s_W B_\mu \; \, ,  \; 
A_\mu = -s_W W_\mu^3 + c_W B_\mu \; \, . 
\end{equation}
Here $s_W  \equiv sin \theta_W$ and  $c_W \equiv cos \theta_W$, where 
$\theta_W$ is the weak mixing angle, which has to 
 be determined by experiment.

The weak currents are
\begin{eqnarray}
j_\mu^{(\pm)} \; = \;  j_\mu^1 \; \pm \; i \, j_\mu^2 \; \, , 
\label{currWtr}
\end{eqnarray}
which  for one electron and its neutrino are obtained in
agreement with (\ref{Fermicur}) and (\ref{curnu}): 
\begin{eqnarray}
j^{(+)}_\mu \, = \,
\overline{\nu_e} \, \gamma_\mu \,L \, e \,  \,
\; \, \, \mbox{and} \; \, \, 
j^{(-)}_\mu \, = \, 
 \overline{e} \, \gamma_\mu \,L \, \nu_e \; \, , 
\label{currWe1}
\end{eqnarray}
where $L=(1-\gamma_5)/2$,

The physical weak and electromagnetic 
currents are known ! For the electron it is
\begin{equation}
j^{em}_\mu \, = \, j^3_\mu \, + \, \frac{1}{2} \, j^Y_\mu
  \; \, = \, - \, \overline{e} \,\gamma_\mu \, e \; ,
\label{jem1} 
\end{equation}
in agreement with (\ref{chargeFormW}).
The neutral current $j^Z_\mu$ {\em times} the coupling $g_Z$ is now 
{\em completely determined}! :
\begin{eqnarray}
j^Z_\mu  \, = \, j^3_\mu  \, - \,  sin^2 \theta_W \, j^{em}_\mu \; ,
\label{currem} 
\end{eqnarray}
where
\begin{equation}
j^{3}_\mu \, = \, \frac{1}{2} \left( \overline{\nu_e} \, 
\gamma_\mu \,L \, \nu_e 
 \, - \,    \overline{e} \, \gamma_\mu \,L \, e \right)  
\label{currZ}
\end{equation}

The couplings for the physical bosons $\gamma, Z, W^\pm$ will be
\begin{equation} 
  g_\gamma \, = g \, sin \theta_W \; , g_Z \, = \, \frac{g}{cos \theta_W} \; , \, g_W \, = \, \frac{g}{\sqrt{2}} \; ,
  \, tan \theta_W \, \equiv  \, \frac{g'}{g} \; \, .
\label{coupl}
\end{equation}
 Experimentally, one finds
 $sin^2 \theta_W \simeq 0.23$ in processes with neutral currents.

 We have obtained a {\it partial unification}: Three couplings have
 become two !

\begin{figure}
\begin{center}
\scalebox{0.4}{\includegraphics{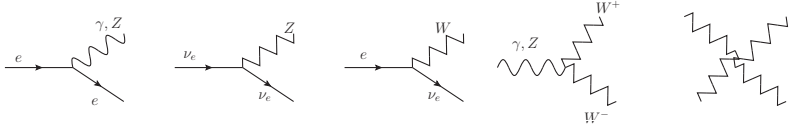}}
\caption{Electroweak vertices for the electron and its neutrino.
The triple and quartic boson vertices are dictated
 by $SU(2)_L \times U(1)_Y$ gauge symmetry. There are also vertices 
with the Higgs boson $H$.}
\label{elweakvert}
\end{center}
\end{figure}

For the two other fermions and their neutrino's there are
are additional terms given
by the same formulae as in this subsection with the replacements
$e \rightarrow \mu$ and  $e \rightarrow \tau$
in eq. (\ref{EWLeptCur}).

\subsection{The Higgs sector}

In 1964, Peter Higgs \cite{Higgs:1964pj}(and others) demonstrated
(in some toy model) how to build
masses into gauge theories without breaking gauge invariance.
One introduces a complex scalar field (and maybe multiplet) $\phi$,
and assumed 
that some part of it had a nonzero value in vacuum, called vacuum
expectation value (VEV). Then this field
could give a mass to the vector(gauge) boson(s).
In a $U(1)$ symmetric theory (like in QED) one could assume an extra
complex field with
 the real part having a nonzero VEV. The imaginary part would then 
correspond to  a massless (Goldstone)
particle  which could be absorbed  and give a mass to
 vector(gauge-boson),  while the rest of the real 
part would be a particle with mass (depending on the value of the VEV). 

The Higgs Lagrangian has the form
\begin{equation}
  {\cal L}_{Higgs} \, = \,
  \left( i D^{(\phi)}_\alpha \, \phi \right)^\dagger \, 
\left( i D_{(\phi)}^\alpha \, \phi \right) \, - \, V(\phi) \; , 
\label{HiggsInt}
\end{equation}
where
\begin{equation}
  V(\phi) \,
\equiv \, \mu^2 \left(\phi^\dagger \, \phi \right) \; +
 \; \lambda \, \left(\phi^\dagger \, \phi \right)^2 \; .
\label{LHiggsV}
\end{equation}
In a simple case  $\phi$ is a complex field, and $D_\mu$ the covariant
derivative as in QED. For $\mu^2$ positive, this would be QED for a scalar
charged particle. Assuming that $\mu^2$ is negative, $\phi$ will get a
VEV $\langle \phi \rangle \, = \, v/\sqrt{2}$ where
$v = \sqrt{-\mu^2/\lambda}$.
Then the vector $A_\alpha$ with a priory two degrees of freedom
will appear with an additional  spin zero component
$\sim e \partial_\alpha (Im(\phi))$ and get a mass $m= e v$.
This mechanism will be demonstrated in the next section for the
$SU(2)_L \times U)1)_Y$ case.

For the $SU(2)_L \times U(1)_Y$ case one introduces a complex SU(2) doublet
\begin{eqnarray}
\phi \, = \,  \left( \begin{array}{c} \phi_+ \\ 
 \phi_0  \end{array}
 \right) \;   ,         
\label{Hdoubl}
\end{eqnarray}
where $\phi$ has a covariant derivative  :
\begin{eqnarray}
i D^{(\phi)}_\alpha \, = \, i \partial_\alpha  \, 
- \, g \, \frac{\tau^k}{2}
 W^k_\alpha - \frac{g'}{2} Y_\phi \, B_\alpha \; \, .
\label{HDeriv}
\end{eqnarray}
  For $\mu^2 >0$,  we have a theory for spin zero particles  with mass $\mu$
  and self-interactions $\sim \lambda$.

  The most simple Lagrangian for interactions between  fermions and
  the complex Higgs doublet field is the $SU(2)_L \times U(1)_Y$
  Yukawa Lagrangian term.
  For the electron (and its neutrino) it reads:
\begin{eqnarray}
{\cal L}_{e\phi} \, = \, - \, G_e \,
 \left( \overline{E_L} \,\phi \, e_R \; 
+ \;  \overline{e_R} \, \phi^\dagger \, E_L \right) \,  .
\label{eYuk}
\end{eqnarray}
This will later be extended with similar terms for the two other
leptons and the quarks. For the $d$-quark (and its partner) it is 
\begin{eqnarray}
{\cal L}_{d \phi} \, = \, - \, G_d
 \, \left( \overline{q^{(1)}_L} \,
 \phi \, d_R \; + 
\;  \overline{d_R} \, \phi^\dagger \,
 q^{(1)}_L \right)   \quad ; \; \,
\overline{q^{(1)}_L} = (\overline{u_L} ,\overline{d_L}) \, .
\label{dYuk}
\end{eqnarray}

Some people say that within the SM, the neutrinos have
 zero mass, by definition.  To me this is a unnecessary definition. 
We  may  perfectly well add  a righthanded neutrino  $\nu_R$ in
the  SM in the same way as the rest of the fermions!
However, it has zero weak hypercharge, and does therefore not couple to
gauge bosons.
It can be introduced  via the adjoint Higgs field $\widetilde{\phi}$
as for the top quark ($t$) and all the other $I_3^w$ particles:
\begin{eqnarray}
 \widetilde{\phi} \; \equiv \phi^c \, = \,  i \tau^2 \, (\phi^\dagger)^T
  \, = \, \left( \begin{array}{c} \phi_0^\dagger \\ 
 - \, \phi_+^\dagger  \end{array}
 \right)  \; .
\label{Hdoubl}
\end{eqnarray}

In general, all couplings $G_f$ must be
determined by experimental values of fermion masses.
To obtain a mass for the neutrino, one might write
\begin{eqnarray}
{\cal L}_{\nu \phi} \, = \, - \, 
 G_\nu \, \left( \overline{E_L} \,\widetilde{\phi} \, \nu_R \; 
+ \;  \overline{\nu_R} \, \widetilde{\phi}^\dagger \, E_L \right)  \; ,
\label{nuYuk}
\end{eqnarray}
and for the top quark:
\begin{eqnarray}
{\cal L}_{t \phi} \, = \, - \, G_t
 \, \left( \overline{q^{(3)}_L} \,
 \widetilde{\phi} \, t_R \; + 
\;  \overline{t_R} \, \widetilde{\phi}^\dagger \,
 q^{(3)}_L \right)   \quad ; \; \,
\overline{q^{(3)}_L} = (\overline{t_L} ,\overline{b_L}) \; .
\label{tYuk}
\end{eqnarray}
There are also ``crossed terms'' which mixes the generations. In subsection E
quark mixing  will be discussed.

{\it If} the $\nu$ is a {\it Majorana particle}, i. e. its own anti-particle,-
this  means that some mechanism
beyond the Standard Model is involved.
Then the neutrino masse may typically be written
according to the ``seesaw mechanism as: 
\begin{equation}
m_v \sim m_l^2/M \; .
\label{seesaw}
\end{equation}
 Here $M$ is the mass of
 some high mass particle (for instance $M \sim 10^?$ GeV), and $m_l$ is a
 lepton mass. \\

$ {\cal N}{\cal B}!$ 
 Up to now we have a completely $SU(2) \times U(1)_Y$ gauge symmetric
 Lagrangian !  The free parameters of this theory are :
\begin{equation}
 g \,\,  , \,  g' \,  \, , \,  \mu^2 \, \, , \, 
\lambda, \mbox{all the} \, \,
  G_i \,  \, \,  \mbox{in} \, \,  {\cal L}_{Yukawa} 
\end{equation}
After symmetry breaking all the physical parameters must depend on these.
The $G_i$'s in the Yukawa's to make fermion masses will later be generalized
to contain the extension in eq. (\ref{YukawaGen}) to obtain quark mixing.
See subsection VIII E.

\subsection{Spontanous Symmetry Breaking (SSB)}

Now we have to break the gauge symmetric theory  in a way such that
the Lagrangian can be intepreted 
in terms of fields for physical particles.

The challenge is now to obtain mass term for the $W$- and $Z$-bosons.
In general,
a  vector field ($V_\mu$) has 
the generic form: 
\begin{equation}
{\cal L}_{Mass} \, \sim  \, m_V^2 \, V_\alpha \, V^\alpha
\end{equation}
Such terms should appear for the $W^\pm$- and $Z$-bosons.

Mass terms for fermions $f$ should also appear in the correct way,
namely as
\begin{equation}
 \sim m_f (\overline{f}_R f_L + \overline{f}_L f_R) .
\end{equation}

Let us now consider the Higgs potential in (\ref{LHiggsV}).
For $\mu^2 > 0$,   the minimum of $V(\phi)$ is at $\phi=0$ , or
 $\langle 0| \phi |0 \rangle =0$ for the quantum case. 
Some people think about the potential that it is like a magnet,
such that at very
high temperatures, the vacuum is symmetrical and
$\mu^2$ is positive. Then, when the temperature is lowered, we may think
that $\mu^2$
becomes negative and the vacuum non-symmetric.
 
If $\mu$ is
{\em assumed} to be  imaginary, and   $\mu^2 < 0$, and at the same time 
$\lambda > 0$, the potential $V(\phi)$ has  minimum for a
value $|\phi| \neq 0$. Then  {\it Spontanous symmetry breaking} (SSB)
will occurr. This means that the  vacuum 
 has not the full symmetry of the dynamical equations.
 The vacuum value of  the field $\phi$ may be  taken to be 
\begin{eqnarray}
\langle 0| \phi |0 \rangle \; = \; \frac{v}{\sqrt{2}} \, \chi_V \; \,
; \; \mbox{where} \; \,   v \, \equiv \sqrt{\frac{-\mu^2}{\lambda}} \; ;
\;
\mbox{and} \, \,  
 \chi_V \, \equiv  
   \;  \left( \begin{array}{c} 0 \\ 
 1  \end{array}
 \right)  \; .
\label{HVac}
\end{eqnarray}
Here the vacuum value $v$ is obtained by taking the derivative of the potential
$V(\phi)$ with respect to $|\phi|$. The form of $\chi_V$ is chosen in order to
obtain masses for the electron, the $W$'s and the $Z$, but not for the photon
(and not for the neutrino at this stage). Note that if one has a
two-component object, one can always find a $2 \times 2$ matrix which will,
by multiplication with the two-component produce a zero at its upper
component. Then the lower component will in general be complex. Then this
complex number may be made real by multiplying it with a suitable
complex number. This explains why the form in (\ref{HVac}) is always possible.
The zero at the top is necessary to obtain thy physics we want (especially
zero photon mass. See later.)
The vacuum value $v$ is measured in GeV. Now the Higgs doublet is split as
\begin{equation}
\phi \, = \, \langle 0| \phi |0 \rangle \; + \; \Delta \phi \; \, ,
\label{phisplit}
\end{equation}
where  $\Delta \phi$ contains two complex fields.
If this is plugged into equation (\ref{HiggsInt})
this form of $\chi_V$ give the SSB.
A triplet $\xi^j(x)$ of 
 massless neutral
(Goldstone) boson
fields and one massive neutral Higgs field $H$ with mass 
$m_H=\sqrt{-2 \mu^2}$ will appear. 

 The Goldstone triplet fields $\xi^i$ ($i=1,2,3$) will be
used as gauge parameters and transformed into three linear
 combinations ($W^{(\pm)}, Z$) of $W^k_\mu$ and $B_\mu$
to {\em make them massive.} This very special $SU(2)_L \times U(1)_Y$ gauge
  transformation contains the Goldstone fields $\xi^j$:
\begin{eqnarray}
\phi \; \rightarrow \phi' \, = \, U_\xi \, \phi \, =  \, \frac{1}{\sqrt{2}}   
  \left( v + H \right) \, \chi_V \; \, . 
\label{HMech}
\end{eqnarray}

This is the ``Physical gauge'',
$H$ is the physical Higgs boson.
The transformation $U_\xi$ has now to be applied to {\it all fields} in our
Lagrangian.
 After Higgs-mechanism has been applied, all transformed  fields depend on 
the Goldstone fields  $\xi^i$.
One sees that the-a priori massless- vector bosons $W^i_\mu$ get
an additional longitudinal
 term  $\partial_\mu \xi^i$ and thereby obtains a mass.

For $Y_\Phi=+1$ (according to $Q_f=T_3^w + Y^w/2$ ),
the transformed covariant derivative on the transformed Higgs doublet
$\phi'$ is now: 
\begin{eqnarray}
\left( i D^{(\phi)}_\alpha \, \phi \right)'
 \; = \; \frac{1}{\sqrt{2}} \,    
     \left( \begin{array}{c}  -g_W \, W^{(-)}_\alpha \,(v + H) \\ 
 i \partial_\alpha \, + \frac{1}{2} \, g_Z \, Z_\alpha \, (v + H)  \end{array}
 \right)  \; \, .                 
\label{Dphitr}
\end{eqnarray} 
${\cal N}{\cal B}!$ The linear combinations for $\gamma$-field $A_\mu$
in  $(D^{(\phi)}_\alpha)'$ {\em has disappeared !} such that $m_\gamma = 0$
Note that now $W_\mu^{(\pm)}$, $Z_\mu$ and $A_\mu$ are now the transformed
fields which are intepreted as the physial ones.
The $U_\xi$ transformed Lagrangian is :
\begin{eqnarray}
{\cal L}_\phi' \, = \, \frac{1}{2}  (\partial_\alpha H) (\partial^\alpha H)
 -   \frac{\mu^2}{2} \, (v + H)^2  -   \frac{\lambda}{4}
 (v + H)^4  \nonumber \\
  + \left(\frac{g}{2} \right)^2  W^{(+)}_\alpha  W^{(-) \alpha} (v + H)^2  
  +  \left(\frac{g_Z}{2}\right)^2  Z_\alpha   Z^\alpha  (v + H)^2 \; \, 
\label{Lphitr}
\nonumber
\end{eqnarray}
 
The physical mass of  Higgs particle  is now
\begin{equation}
m_H = \sqrt{\mu^2 + 3 \lambda v^2} \, = \,  \sqrt{-2 \mu^2} \, .
\end{equation}

The photon $\gamma$ remains massless, while 
 the masses of $W$ and $Z$-bosons are given by 
\begin{equation}
M_W = \frac{1}{2} g \, v \; \, , M_Z \, = \, \frac{1}{2} \, g_Z \, v \; , \,
\Rightarrow \frac{M_Z}{M_W} \, = \frac{1}{cos \theta_W} \; \, . 
\label{Vmasses}
\end{equation}
    {\it This relation for the ratio between the $Z$ and $W$ masses
      must hold experimentally!!!}
(-up to higher corrections)\\

 From $\alpha_{em}$ and $G_F$, one finds 
$v \simeq $ 246 GeV.
Gauge invariance implies some relations between couplings and  masses.
If some of these relations are broken, then gauge invariance will also
be broken.

\subsection{Quarks in electroweak interactions}

In 1967, - for people who believed in quarks-,  only three quark types
 were known. The lightest, $u$ and $d$
could be put in one doublet. But what about  an eventual second doublet ?
The $s$-quark was established through the existence of the $\Sigma$ particles.
But the $s$-quark had at this time no partner, in contrast to  the $u$- and
$d$-quarks.

In order to explain the beta-decay 
of the $\Sigma^0$ particle, 
 $\Sigma^0 \rightarrow p \, e^- \, \overline{\nu_e} \, ,$
the italian physcisist Cabibbo \cite{Cabibbo:1963yz} introduced a mixing of $d$ and $s$-quarks
which have the same electric charge. This  
Cabibbo mixing is :
\begin{equation}
 d \rightarrow d_\theta = d \, cos \theta \, + s \, sin \theta \, , \,
 \mbox{and}  \;  s \rightarrow s_\theta = s \, cos \theta \, -
 d \, sin \theta \; .
\label{Cabibbo}
\end{equation}
Here $\theta \equiv \theta_C$  is the  Cabibbo mixing angle. The ratio of
$\Sigma^0$ and $n$ beta 
decay amplitudes should be  $ = \, tan \theta_C$, in the limit of equal neutron and $\Sigma^0$ masses. Experimentally, 
one finds $sin \theta_C \simeq 0.225$ from comparing the beta-decays of $n$ and $\Sigma^0$. 

\begin{figure}
\begin{center}
\scalebox{0.45}{\includegraphics{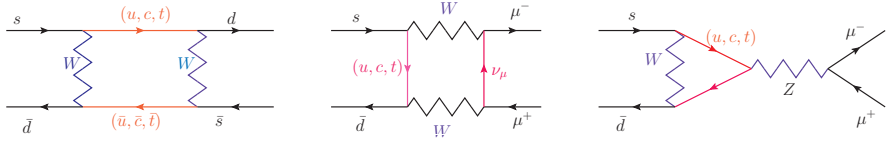}}
\caption{Quark diagram for $\overline{K^0} \rightarrow K^0$ determiing the
  mass difference 
  $\Delta m_K=(m_L-m_S)$ and  the CP-violating parameter
  $\epsilon$ in (\ref{KLstate}) (left diagram), and diagrams for
 $\overline{K^0} \rightarrow \mu^+ \mu^-$. It was found from these processes 
that $m_c < $ 2 GeV/$c^2$  }
\label{K-decays}
\end{center} 
\end{figure}

Now the $d_\theta$ was put in a doublet with the $u$-quark, but the
field $s_\theta$ had no partner, so 
 there was one quark type ``missing''. However, already in 1964
Bjorken had suggested the existence of
a fourth quark $c$ with the same charge as the $u$-quark.
More important,
in 1970, Glashow, Iliopoulos and Maiani \cite{Glashow:1970gm}
(named {\it GIM}), within the
$SU(2)_L \times U(1)_Y$
scheme, 
postulated the  existence of fourth quark $c \,$ (={\it charm})
in order to explain the mass difference $\Delta m_K$ of the particles
$K_L$ and  $K_S$ in (\ref{Kstates}) and (\ref{KLstate}). 
From the diagrams in Fig. \ref{K-decays}, with $c$- and $u$-quarks in
the loop, one found that the $c$-quark
should have a mass less than 2 GeV/$c^2$.
Now there was  two quark doublets: 
\begin{equation}
 \left( \begin{array}{c} u \\ d_\theta  \end{array} \right)_L 
\;  , \;  
 \left( \begin{array}{c} c \\ s_\theta  \end{array} \right)_L \; \, , 
\end{equation}
and four singlets  $u_R , (d_\theta)_R , (s_\theta)_R, c_R$.
Because of the orthogonality of the quark mixing in (\ref{Cabibbo}),
there were partial cancellations between $u$ and $c$-quark contributions for
the process $s \overline{d} \rightarrow \mu^+ \, \mu^-$ as for
in $s \overline{d} \rightarrow d \overline{s}$ (the $K_L - K_S$
mass difference) in  Fig. \ref{K-decays} .

The decays of $K$-mesons played an important role in determining
the structure of weak interactions. This  was an example of what
is often called {\it flavor physics}.

Later, a new quark called $b$-quark (bottom quark) was found, and a sixth 
quark $t$ (top-quark)  was introduced, anticipated, in order to fulfil
a third family. Then the idea of Cabibbo in (\ref{Cabibbo})
was generalized by Kobayashi and Maskawa \cite{Kobayashi:1973fv}, and
for   three families, {\it i.e}
three lepton doublets and three quark doublets we have:
\begin{equation}
 \left( \begin{array}{c} \nu_e \\ e^-  \end{array} \right)_L 
 \, ,  \,
 \left( \begin{array}{c} \nu_\mu \\ \mu^-  \end{array} \right)_L  
\, , \,
 \left( \begin{array}{c} \nu_\tau \\ \tau^-  \end{array} \right)_L  
 \left( \begin{array}{c} u \\ d_{CKM } \end{array} \right)_L 
\, , \,
 \left( \begin{array}{c} c \\ s_{CKM}  \end{array} \right)_L  
\, , \,
 \left( \begin{array}{c} t \\ b_{CKM}  \end{array} \right)_L \; \, ,
\end{equation}
-plus all the right-handed singlets.  Now all the quarks with charge -1/3 mix,
 and we have the CKM (Cabibbo-Kobayashi-Maskawa)- mixing of quarks
\begin{eqnarray}
 d_{CKM} = V_{ud} \, d + \, V_{us} \, s + \, V_{ub} \, b \; ,
\nonumber \\
 s_{CKM} = V_{cd} \, d + \, V_{cs} \, s + \, V_{cb} \, b \; ,
\nonumber \\
 b_{CKM} = V_{td} \, d + \, V_{ts} \, s + \, V_{tb} \, b \; ,
\label{CKMterms}
\end{eqnarray}
which is a generalsitation of (\ref{Cabibbo}).
Here $(d,s,b)$ are the physical quarks (with definite mass), and 
$V_{ij}$'s are Cabibbo-Kobayashi-Maskawa matrix elements.

This kind of quark mixing is forced to occurr for the following reason:
There is  no symmetry which can prevent mass terms of the
type $\overline{c_R} \, u_L$
or $\overline{s_R} \, d_L$, say. But particles have a definite mass.
Therefore diagonalisation of mass matrices must be performed.
And the mismatch between diagonalisation in the charcge +2/3 sector
and -1/3 sector is just the $V_{CKM}$ matrix. 
This is shown explicitly in the next subsection (E).

Decays of kaons ($K$-mesons) have played an important role in 
understanding of the SM. Some of the elements of $V_{CKM}$ are
complex! This means that there is   
$CP$-violation in the SM. Within this picture, the  $CP$-violation
observed in the decays of $K$-mesons is explained !
(-at least the electroweak part. The non-perturbative QCD part
to get the (more) precise number is worse!)

Now neutrinos are known to have tiny but non-zero masses. This means
that there is  a  mixing also in the neutrino sector. If this mixing
has the same 
origin as the other fermions or if it as another like for instance of
Majorana type is not clear.

\subsection{Formal description of quark mixing}

In general, Lagrangian terms  generating fermion masses are generalizations of
the terms in  (\ref{dYuk}) and (\ref{tYuk}):
 $\left( \sim m (\overline{\psi}_R \psi_L + \overline{\psi}_L \psi_R) \right)$:
\begin{eqnarray}
{\cal L}_{f \phi} \, = \, - \, \sum_{f_L, f_R} \,
 C_{f_L f_R} \, \left( \overline{f_L} \,\phi \, f_R \;
+ \;  \overline{f_R} \, \phi^\dagger \, f_L \right) \,
\nonumber \\
 + \, \sum_{f_L, \widetilde{f_R}}
 \widetilde{C}_{f_L \widetilde{f_R}} \, \left( \overline{f_L} \,
 \widetilde{\phi} \, \widetilde{f_R} \; +
 \;  \overline{\widetilde{f_R}} \, \widetilde{\phi}^\dagger \,
 f_L \right) \Big] \; .
\label{YukawaGen}   
\end{eqnarray}
Note: There are two right-handed singlets ($f_R, \widetilde{f_R}$)
per left-handed doublet ($f_L$).
With such Lagrangian terms one obtains diagonal and non-diagonal
mass terms in the quark sector.
After SSB there will  be
non-diagonal  mass-terms  like
like $\bar{u_L'} \, c_R' \;$ (charge 2/3 quarks), $\bar{d_L'} \, b_R'$
(charge -1/3 quarks ). Such terms are  not forbidden by
the symmetry of the $SU(2)_L \times U(1)_Y$, and we obtain a mass part of
the Lagrangian which have this form:
\begin{eqnarray}
{\cal L}_{mass}(q)'  \, = \, - \, \overline{ {\cal U}_W} 
\, {\cal M}_{{\cal U}} \, {\cal U}_W 
 \, - \, \overline{ {\cal D}_W} 
\, {\cal M}_{{\cal D}} \, {\cal D}_W \; ;
\label{LMass}
\end{eqnarray}
\begin{eqnarray}
 {\cal U}_W \, \equiv \, \left( \begin{array}{c} u_W \\
 c_W \\ t_W  \end{array} \right) \;\; ; \;  
 {\cal D}_W \, \equiv \,  \left( \begin{array}{c} d_W \\
 s_W \\ b_W  \end{array} \right) \; ,
\label{Wstates}
\end{eqnarray}
where the fields with subscript $W$ are the transformed (primed) states,
also called the weak eigenstates. The mass matrices ${\cal M}_{{\cal U}}$ and ${\cal M}_{{\cal D}}$ have 
diagonal and non-dagonal mass matrix elements which are proportional to
$v$ and the coefficients $C$ in ${\cal L}_{f \phi}$.

Physical fermion  states (mass eigenstates) have diagonal
mass matrices. Diagonalization of the mass matrices is performed by a
$3 \times 3$ matrix $S_{{\cal Q}}$ goes like this:
\begin{eqnarray}
 {\cal M}_{\cal Q}^{diag} \, = \,  S_{{\cal Q}} \, {\cal M}_{\cal Q} \,
 S_{{\cal Q}}^\dagger \qquad ; \quad  {\cal Q}_P =  S_{{\cal Q}} {\cal Q}_W
\label{MassDiag}
\end{eqnarray}
for ${\cal Q}={\cal U}$ and  ${\cal Q}={\cal D}$, respectively.
 The physical quark states are (the triplets of) {\em mass eigenstates}: 
 ${\cal U}_P \,$, ${\cal D}_P$.

The weak currents  for three generation of quarks are:
\begin{eqnarray}
j_\mu^{(+)}(q)  \, = \, \overline{ {\cal U}_W} \; \gamma_\mu \, L \,
 {\cal D}_W  \; = \;
 \overline{ {\cal U}_P} \; \gamma_\mu \, L \,  {\cal D}_{CKM} \; ,
\end{eqnarray}
\begin{eqnarray}
j_\mu^{(-)}(q)  \, = 
\, \overline{ {\cal D}_W} \, \gamma_\mu \, L \, {\cal U}_W \, =
 \, \overline{ {\cal D}_P} \; \gamma_\mu \, L \, V_{CKM}^\dagger
\; {\cal U}_P \; ,
\label{GIMKM-}
\end{eqnarray}
 \begin{eqnarray}
 {\cal D}_{CKM} \; \equiv   \, V_{CKM} \,  {\cal D}_P \quad  ; \; 
 {\cal U}_{CKM}  \; \equiv   \,  V_{CKM}^\dagger \; {\cal U}_P \; , 
\label{GIMKM+}
\end{eqnarray}
 where
\begin{eqnarray}
 V_{CKM} \equiv  S_{{\cal U}} \, S_{{\cal D}}^\dagger
 \label{KobMask}
\end{eqnarray}
is the Kobayashi-Maskawa quark mixing matrix.

 Note that $ S_{{\cal U}} \neq S_{{\cal D}}$ implies $V_{CKM} \neq 1$
 (Later we might drop  subscript $P$).
 That is, the quark mixing matrix $V_{KM}$ 
 the is the result of the mismatch of  diagonalization in the up
 ($U$-type)
 vs. the down ($D$-type) sector of quarks. Then we have explained
 why the relations appear. We also see that we might also have put
 the mixing in the up-quark sector, {\it i.e} among the $(u,c,t)$ quarks.
 But for historical reason, to generalise Cabibbo mixing, it is put in the
 down-quark sector $(d,s,b)$.
 The quark mixing matrix $V_{KM}$
 is a unitary matrix with three real and one imaginary independent
 parameters. These have to 
 be determined experimentally.
 One might use three angles and one phase factor. The mixed parts
 $(d_{CKM},s_{CKM},b_{CKM})$ of ${\cal D}_{CKM}$ are given in the
 equations  in  (\ref{CKMterms}). The fields $(d,s.b)$ are those
 in ${\cal D}_{P}$.

 Quark mixing means that quark flavor is not conserved! 
 The heavier
quarks will decay to lighter quarks. In terms of hadronic particles it means
that for instance $B$-mesons might decay to $D$-mesons (and lighter mesons),
$D$-mesons to $K$-mesons, which decays to pions.
The Imaginary part of $Im(V_{q q'}) \neq 0$. This
{\em breaks $CP$-invariance} ! -
for instance in: $K \rightarrow 2 \pi$ and $B \rightarrow \psi K_S$ decays.

However: It is found that
{\it  $CP$-violating effects in the $SM$ are too small to
  explain early universe cosmology}.

\begin{figure}
\begin{center}
\scalebox{0.5}{\includegraphics{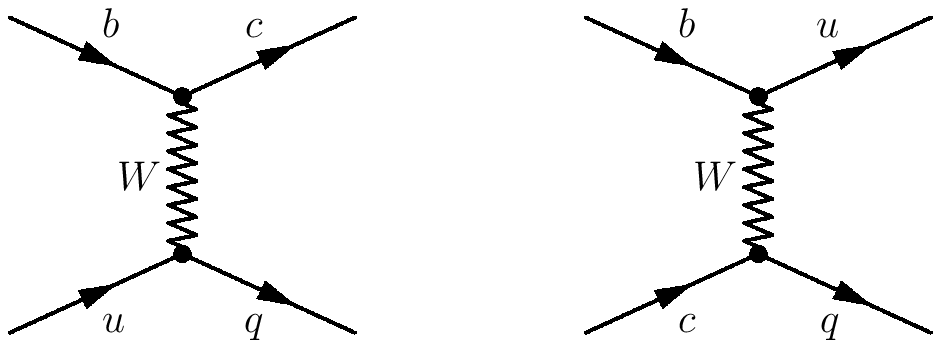}}
\caption{Basic W-exchange diagrams generating various  non-leptonic decays
  of $B$-mesons. (Here $q=d$ or $s$). Some quark lines have to be bended,
  and a ``spectator quark'' line has to be added in order to obtain a
  diagram for a physical process.}
 \label{fig:4QWint}
\end{center}
\end{figure}

For the  neutral currents  $j_\mu^{em} \; , \; j_\mu^{Z}$ 
the quark mixing  (flavor changing) disappear ({\em GIM-mechanism}) because 
$S_{{\cal Q}}^\dagger \, S_{{\cal Q}} =1$ leads (for
 ${\cal Q}={\cal  U}, {\cal D}$)  to  
 $ \overline{{\cal Q}_W} \, \Gamma \, {\cal Q}_W \, = \, 
\overline{{\cal Q_P}} \, \Gamma \, {\cal Q_P} $,
 where $\Gamma$ is some (product of) Dirac matrices.

 In the quark sector the neutral currents are :
\begin{eqnarray}
  j_\mu^{em}(q)  \, = \, \frac{2}{3} \,
  \overline{ {\cal U}} \; \gamma_\mu \, {\cal U} \, 
-  \frac{1}{3} \, \overline{ {\cal D}} \; \gamma_\mu  \, {\cal D}  \; ,
 \\
j_\mu^{3}(q)  \, = \, \frac{1}{2} \left(\overline{ {\cal U}} \;
\gamma_\mu \, L \, {\cal U} \, 
-   \, \overline{ {\cal D}} \; \gamma_\mu  \, L  {\cal D} \right)  \; .
\label{NeutrC}
\end{eqnarray}
These expressions are valid both for 
${\cal Q}_P$ and  ${\cal Q}_W$ because 
{\it the neutral currents are diagonal in the quark fields}.
The flavor changing neutral current
(FCNC) processes occurr only at loop level (examples
$b \rightarrow s \gamma$, for instance  in $B\rightarrow \gamma K^*)$.

\section{Summing up the Standard Model(SM)}

\subsection{Full perturbative $SU(3)_c \times SU(2)_L \times U(1)_Y$}

\begin{center}
\begin{figure}[htbp]
 \scalebox{0.4}{\includegraphics{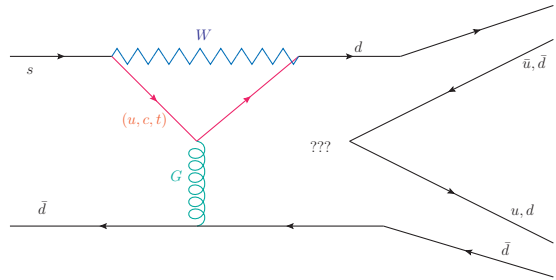}}
\caption{Example: ``Penguin diagram''.Perturbative weak and strong 
interactions in play in a triangle loop(quantum fluctuation). AND we se a
Non-perturbative QCD interaction symbolized by the ??-sign .}
\label{kto2pi}
\end{figure}
 \end{center}

In many processes, the full SM is in play, both electroweak interactions
as well as QCD, both perturbative and non-perturbative regions. An example
is shown in Fig. \ref{kto2pi}. 
This diagram  illustrates the proesses
$K_L \rightarrow \pi^+ \pi^-$ -or $2 \pi^0$. 
$CP$-violation is different for charged and neutral pions 
(This is the so called  $\epsilon'$-effect)
The non-perturbative QCD part is difficult. It has been handeled by
{\it lattice gauge theory}. Here the initial $s$ and $\bar{d}$ quarks are bound
(by non-perturbative QCD) to a $\overline{K^0}$. The non-perturbative
forming of $\bar{d} d$ to two pions is merked by ``??''. In the physics
of kaon decays, 
the penguin-like interactions which are loop diagrams for
$s \rightarrow d \, g$, $s \rightarrow d \, g g$, $s \rightarrow d \gamma$,
$s \rightarrow d \, g \gamma$, $s \rightarrow d \, \gamma \gamma,$ and so on,
has played an important role. I worked with such loops for many years
myself \cite{Eeg:1988td}.

\subsection{The Higgs. The final building stone}

\begin{figure}
\begin{center}
\scalebox{0.30}{\includegraphics{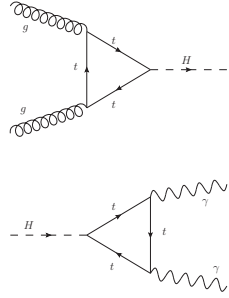}}
\caption{Feynman diagram for Higgs production through gluon fusion, via a
  triangle top quark. Remember that the Higgs couples stronger to heavier
  particles. Then the  Higgs might deay to two photons.
  }
\label{Higgs-diagr}
\end{center}
\end{figure}

NB! : A proton at high energy contains gluons, as mentioned before. 
When two protons collide, a gluon from one of the protons might collide
 with a gluon from the other, {\it i.e.} two gluons might collide.
Then  the Higgs particle might appear as a 
fusion  from two  gluons and make, via a top-quark loop, a Higgs boson.
But the Higgs boson is
very unstable and may decay into two photons. See Fig. \ref{Higgs-diagr}.

\subsection{ Essentials of the SM}
The Standard Model(SM) is formulated within
quantum field theories(QFT).
One might say that
the  particles, leptons and quarks and 
their  antiparticles, and also the gauge bosons,
 are exitations of their corresponding fields.
In QFT the number of particles are not
 conserved, only the total
 energy and the (electric) charge.
 Particles might appear, live for a 
very short time  and disappear, as {\it quantum fluctuations}.
I this sense, QFT goes beyond quantum mechanics.
In QFT gauge invariance has been a ``guiding principle''.
 Gauge-symmetric theories have been chosen because they
 are {\em renormalizable}. That means that calculations can be
 performed in a controllable, consistent way.
 Gauge-symmetric theories are also chosen because
 they have a minimum of parameters, and of course
 because it is in {\it agreement with experiment} with these minimal
 number of parameters.

 Looking at the electroweak $SU(2)_L \times U(1)_Y$ part first:
 In the manifest symmetric version the free parameters are the coupligs
 $g$ and $g'$, the parameters $\mu$ and $\lambda$ in the Higgs potential,
 and all the  parameters of type $C_{f f'}$ in the general Yukawa
 interaction in (\ref{YukawaGen}), which later make fermion masses and
 mixings.

 After the spontanous symmetry breaking(SSB and Higgs-mechanism) of
 $SU(2)_L \times U(1)_Y$, the
free parameters of the electroweak part are
the electric charge $g_\gamma$ \, $\theta_W \,$, all 12 fermion
masses $m_f$,
the four independent paramameters of  quark mixing
matrix $V_{KM}$. If the neutrino-mixing are of the same origin as in
the quark sector, one should count also these. The Higgs
mass $M_H = \sqrt{-2 \mu^2}$ and  Higgs self-interacting
coupling $\lambda $ are also free parameters (to be determined by
experiment).
Their ratio is fixed by $v=\sqrt{-\mu^2/\lambda}$. Thus, if $M_H$ and
$\lambda$ is known, then $v$ and thereby $G_F$ is known.
Historically, it is the other way around. $G_F$ was known, and
thereby $v$ is known.
Then if $M_H$ is measured, $\lambda$ will be known from the values of $M_H$
and $v$.  BUT as a side remark: It should be noted that the Higgs
mechanism would work also
if the Higgs potential has a slightly different shape and could contain
more parameters. And more general, maybe the model in
equation (\ref{HiggsInt}) is an effective theory mimicing some deeper theory.  

 Many relations between  couplings of bosons to fermions, triple and
 quartic couplings of vector bosons are fixed by gauge symmetry.
 
For QCD the coupling $g_s$ is a free parameter. Here the analogue of
fine structre
$\alpha_{em}$, namely $\alpha_s \, = \, g_s^2/(4\pi)$ is measured at
some energy (chosen as $M_Z c^2$). Then loop calculations and renormalization
theory will give us the result for other energies. 
There is also the parameter $\theta_{QCD}$ for the QCD anomaly which I
have not talked about.
 
Perturbative QCD breaks down at low energies.
  Lattice gauge theory may/should solve the problem.
And it have up to now also solved some problems.  
Going to non-parturbative QCD there are parameters
like $f_\pi$ in $\chi PT$ and further all the masses of the mesons and
baryons....

 So far we can conclude that the SM is valid- as far as we can measure.

\subsection{Beyond the SM?}

It is now clear that the masses of neutrinos are very small but non-zero.
Thus there are a mixing in the leptonic sector,- similar to the
mixing  in
the quark sector- This implies that
lepton flavors $L_e \, , L_\mu \, , L_\tau $
are not separately conserved.
We have {\em neutrino oscillations}. Neutrinos
may change flavor when they travel from the sun to us.
And processes like 
 $\mu^\pm    \rightarrow e^\pm \gamma$  are not forbidden.
However, the origin of the neutrino masses are not established.
Maybe some physics beyond the SM is involved.

Through the decades the SM has existed, many models beyond the SM (BSM)
has been proposed, and some more or less falsified, or at least do not
seem to be so attractive any more.
Among the most important ideas are: {\it Grand Unification (GUT)
 and Supersymmetry(SUSY)}. GUTs tries to unify all the interactions in the SM.
The simplest $SU(5)$ version of GUT could predict
$(sin_{\theta_W})^2$ and a decay of the proton.
The value for $(sin_{\theta_W})^2$  were relatively close to the experimental
value, but still not close enough. Also no proton decay is seen so far.

SUSY implies that every particle has a Heavy SUSY partner with a
different spin.
Fermions have bosonic partners and bosons have fermionic partners.
It has been suggested that adding SUSY and going to bigger GUT gauge
groups might be favorable. But there
is still no conclusion.  And no SUSY particles have  been seen.
The general conclusion is  that still no
``New Physics'', {\it i.e. BSM, is found yet}, but it is
still {\it not excluded}. 

\hspace{4cm} * \hspace{2cm}  * \hspace{2cm} *

This article is partly based on lectures hold at the ``Oslo winter school''
managed by Larissa Bravina and coworkers in 2018 and 2020. It is also partly
inspired by regular lectures I gave at the university of Oslo in the fall 2021.

\bibliographystyle{unsrt}

\end{document}